\documentclass[12pt]{article}
\usepackage{amsmath,amsfonts, amssymb, braket, bbold}
\usepackage{tikz}
\usetikzlibrary{trees,er,snakes,shapes,mindmap}
\usepackage{titlesec}
\titleformat{\section}
 {\normalfont\fontfamily{put}\fontsize{12pt}{16pt}\bfseries\color{black}}
{\thesection}{1em}{}
\titleformat{\subsection}
 {\normalfont\fontfamily{put}\fontsize{12pt}{16pt}\bfseries\color{black}}
{\thesubsection}{1em}{}
\def \beq  {\begin{equation}}
\def \eeq  {\end{equation}}
\def \beqar {\begin{eqnarray}}
\def \eeqar {\end{eqnarray}}
\allowdisplaybreaks
\def\sqr#1#2{{\vcenter{\vbox{\hrule height.#2pt
\hbox{\vrule width.#2pt height#1pt \kern#1pt
\vrule width.#2pt}\hrule height.#2pt}}}}

\def\la {{\langle}}
\def\ra {{\rangle}}

\def\vf {{\varphi}}

\def\Tr {{\rm Tr}}

\def\ba {\bar{a}}
\def\bD {\bar{D}}
\def\bA {\bar{A}}

\def\bx {\bar{x}}
\def\by {\bar{y}}

\def\bnabla {\bar{\nabla}}

\def\del {\partial}
\def\bdel{\bar{\partial}}
\def\a {\alpha}
\def\b {\beta}
\def\e {\epsilon}

\def\bz {{\bar{z}}}

\def\A {{\cal A}}

\def\C {{\cal C}}
\def\D {{\cal D}}

\def\G {{\cal G}}

\def\M{{\cal M}}

\def\P {{\cal P}}

\def\vf {{\varphi}}

\def\half{\textstyle{1\over 2}}

\mathchardef\mhyphen="2D
\begin{document}
%%%%%%%%%%%%%%%%%%%%%%%%%%%%%%%%%%%%%%%%%%%%%%%
%%%%%%%%%%%%%%%%%%%%%%%%%%%%%%%%%%%%%%%%%%%%%%%
\fontfamily{bch}\fontsize{12pt}{16pt}\selectfont
%\fontfamily{pnb}\fontsize{12pt}{16pt}\selectfont
%\fontfamily{pzc}\fontsize{14pt}{16pt}\selectfont
%\fontfamily{pbk}\fontsize{12pt}{16pt}\selectfont
%\fontfamily{cmr}\fontsize{11pt}{15pt}\selectfont
%\fontfamily{put}\fontsize{12pt}{17pt}\selectfont
%\fontfamily{lmss}\fontsize{11pt}{16pt}\selectfont
%\fontfamily{phv}\fontshape{ro}\fontsize{11pt}{14pt}\selectfont
%\fontfamily{ptm}\fontseries{m}\fontshape{r}\fontsize{12pt}{16pt}\selectfont
%\fontfamily{pnc}\fontseries{m}\fontshape{r}\fontsize{11pt}{15pt}\selectfont
%\fontfamily{ppl}\fontseries{m}\fontshape{r}\fontsize{11pt}{15pt}\selectfont
%\usefont{T1}{phv}{m}{it}
%%%%%%%%%%%%%%%%%%%%%%%%%%%%%%%%%%%%%%%%%%%%%%%
%%%%%%%%%%%%%%%%%%%%%%%%%%%%%%%%%%%%%%%%%%%%%%%
\def \CMP {{Commun. Math. Phys.}}
\def \PRL {{Phys. Rev. Lett.}}
\def \PL {{Phys. Lett.}}
\def \NPBProc {{Nucl. Phys. B (Proc. Suppl.)}}
\def \NP {{Nucl. Phys.}}
\def \RMP {{Rev. Mod. Phys.}}
\def \JGP {{J. Geom. Phys.}}
\def \CQG {{Class. Quant. Grav.}}
\def \MPL {{Mod. Phys. Lett.}}
\def \IJMP {{ Int. J. Mod. Phys.}}
\def \JHEP {{JHEP}}
\def \PR {{Phys. Rev.}}
\def \JMP {{J. Math. Phys.}}
\def \GRG{{Gen. Rel. Grav.}}
%%%%%%%%%%%%%%%%%%%%%%%%%%%%%%%%%%%%%%%%%%%%%%%
%%%%%%%%%%%%%%%%%%%%%%%%%%%%%%%%%%%%%%%%%%%%%%%
\begin{titlepage}
\null\vspace{-62pt} \pagestyle{empty}
\begin{center}
%\rightline{CCNY-HEP-18/4}
%\rightline{August 2018}
\vspace{1.3truein} {\large\bfseries
Gauge and Scalar Fields on $\mathbb{CP}^2$: A Gauge-invariant Analysis }
\vskip .2in
{\large\bfseries II. The measure for gauge fields and a 4d WZW theory}\\
~\\
%%%%%%%%%%%%%%%%%%%%%%%%%%%%%%%%%%%%%%%%%%%%%%%
%%%%%%%%%%%%%%%%%%%%%%%%%%%%%%%%%%%%%%%%%%%%%%%
{\sc Dimitra Karabali$^{a,c}$, Antonina Maj$^{a,b,c}$, V.P. Nair$^{b, c}$}\\
\vskip .2in
{\sl $^a$Physics and Astronomy Department,
Lehman College, CUNY\\
Bronx, NY 10468}\\
\vskip.1in
{\sl $^b$Physics Department,
City College of New York, CUNY\\
New York, NY 10031}\\
\vskip.1in
{\sl $^c$The Graduate Center, CUNY\\
New York, NY 10016}\\
 \vskip .1in
\begin{tabular}{r l}
{\sl E-mail}:&\!\!\!{\fontfamily{cmtt}\fontsize{11pt}{15pt}\selectfont 
dimitra.karabali@lehman.cuny.edu}\\
&\!\!\!{\fontfamily{cmtt}\fontsize{11pt}{15pt}\selectfont amaj@gradcenter.cuny.edu}\\
&\!\!\!{\fontfamily{cmtt}\fontsize{11pt}{15pt}\selectfont vpnair@ccny.cuny.edu}\\
\end{tabular}
\vskip .5in

%%%%%%%%%%%%%%%%%%%%%%%%%%%%%%%%%%%%%%%%%%%%%%%
%%%%%%%%%%%%%%%%%%%%%%%%%%%%%%%%%%%%%%%%%%%%%%%
\centerline{\large\bf Abstract}
\end{center}
We consider the volume of the gauge orbit space for
gauge fields on four-dimensional complex projective space.
The analysis uses a
parametrization of gauge fields where gauge transformations act homogeneously on the fields, facilitating a manifestly gauge-invariant analysis.
The volume element contains a four-dimensional
 Wess-Zumino-Witten (WZW) action for a hermitian matrix-valued
 field. There is also a mass-like term for certain components of the
 gauge field.
We discuss how the mass term could be related to results
from lattice simulations as well as Schwinger-Dyson equations.
We argue for a kinematic regime where the Yang-Mills theory can be approximated by the 4d-WZW theory.
The result is suggestive of 
the instanton liquid picture of QCD. Further it is also indicative
of
the mechanism for confinement
being similar to what happens in two dimensions.

\end{titlepage}
%%%%%%%%%%%%%%%%%%%%%%%%%%%%%%%%%%%%%%%%%%%%%%%
%%%%%%%%%%%%%%%%%%%%%%%%%%%%%%%%%%%%%%%%%%%%%%%
\fontfamily{bch}\fontsize{12pt}{17pt}\selectfont
\pagestyle{plain} \setcounter{page}{2}
\section{Introduction}

In this paper, we continue our analysis of quantum fields on the manifold
${\mathbb{CP}^2}$ focusing on the gauge-invariant volume element relevant to the functional integration over gauge fields \cite{KMN1}.
As is well known, in a gauge theory, the
physical degrees of freedom correspond to the space of all
gauge potentials ($\A$) modulo the set of all
gauge transformations which are set to the identity at some chosen point of the spacetime manifold ($\G_*$).
The volume element on the gauge orbit space $\C = \A/\G_*$ is what 
appears in the functional formulation of gauge theories, eliminating the redundant degrees of freedom corresponding to gauge transformations.
Despite being a key foundational ingredient for the 
quantum description of gauge theories, there is no satisfactory expression for this volume element for four-dimensional nonabelian gauge theories
\cite{singer}.
While perturbation theory is well understood by use of gauge-fixing and Faddeev-Popov ghosts, or, equivalently, via the BRST (Becchi-Rouet-Stora-Tyutin) procedure, an analytic approach to nonperturbative questions such as confinement remains elusive.

The motivation to consider nonabelian gauge theories
on the manifold ${\mathbb{CP}^2}$
is from two and three dimensions. In two dimensions, it is possible to calculate the volume element for $\C$ exactly in terms of a Wess-Zumino-Witten (WZW) action \cite{gawe}.
The same calculation placed within a Hamiltonian formulation of
(2+1)-dimensional gauge theories has led to an analytic formula for the string tension \cite{{KKN},{nair-trento1}} which agrees very well, to within about 2\%, with the results from numerical simulations \cite{teper}. Another prediction regarding the Casimir effect for nonabelian gauge theories also seems to agree, to within one percent or so, 
with numerical estimates \cite{chernodub}. The key feature which facilitated such calculations was the complex structure of the two-dimensional spaces and an associated parametrization of gauge potentials which allowed for factoring out gauge transformations and reduction to gauge-invariant degrees of freedom in a simple way. The manifold $\mathbb{CP}^2$ then emerges as the natural candidate for a similar scenario in four dimensions.
This is a complex K\"ahler manifold with a standard choice of metric as the Fubini-Study metric which is given in local complex coordinates
$z^a$, $\bz^{\bar a}$, $a= 1,2$, $\ba = 1, 2$, by
\beq
ds^2 = {dz \cdot d\bz \over (1+ z\cdot \bz/r^2)}
- {\bz \cdot dz \, z\cdot d\bz \over r^2 (1+ z\cdot \bz /r^2)^2}
= g_{a {\bar a}} dz^a d\bz^{\bar a}
\label{cpII-1}
\eeq
Here $r$ is a scale parameter defining the volume of the space as
$\pi^2 r^4 /2$. As $r \rightarrow \infty$, the metric becomes flat
(albeit modulo some global issues), 
so that one can compare results with expectations in flat space.
The main advantage is that $\mathbb{CP}^2 = SU(3)/U(2)$, so that, utilizing group theoretic techniques,
one can obtain a parametrization of gauge fields similar to what was obtained in two dimensions. This was indicated in \cite{nair} and explained in detail
in \cite{KMN1}. In the latter paper, we calculated the leading
quantum corrections, i.e., monomials of fields and derivatives of the lowest dimensions generated by loops, due to a chiral scalar field on $\mathbb{CP}^2$ coupled to gauge fields.
The effective action from integrating out the scalar fields comprised of a quadratic divergence corresponding to a possible gauge-invariant mass term, standard logarithmic divergences corresponding to wave function and/or coupling constant renormalization and a finite
WZW action, which is a dimensionally upgraded version of the 2d-WZW action.
The natural next set of questions will be about contributions due to the gauge fields themselves, with the volume element on $\C$ being a key part of the one-loop
results. This is the subject of the present paper.

The organization of the paper is as follows. The parametrization of the gauge fields and the factoring out of the gauge degrees of freedom are reviewed
in section 2. The formal expression for the volume element for the gauge orbit space $\C = \A /\G_*$ is given in section 3, where we also
identify the relevant Jacobian determinant to be calculated.
In addition to the scalar propagator on $\mathbb{CP}^2$
(with hypercharge $Y = 0$), which was calculated in \cite{KMN1},
we will need the propagator for an antisymmetric rank-2 tensor
(with $Y = -2$). This is calculated in section 4.
A covariant point-splitting regularization, consistent with the isometries 
of $\mathbb{CP}^2$,
is discussed in section 5.
In section 6, we give the key results of our calculations.
We have calculated the terms of the lowest scaling dimension
($\leq 4$), which are presumably the most relevant for
the long wavelength modes of the fields.
These include a four-dimensional WZW term with a finite coefficient,
a mass-like term with a quadratically divergent coefficient and
a set of log-divergent terms of dimension $4$.
The physical implications of these results are discussed in section 7.
There are two appendixes which give details of the calculations for
the WZW term and the UV divergent terms. 

\section{Parametrization of gauge fields}

We start by recalling the parametrization of the gauge fields introduced in
\cite{KMN1}.
The manifold $\mathbb{CP}^2$ is taken to be the group coset space
$\mathbb{CP}^2 = SU(3)/ U(2)$, so that
it can be 
coordinatized in terms of a group
element $g \in SU(3)$, with identification
$g\sim  g h$, $h \in U(2) \subset SU(3)$.
Thus $U(2) $ defines the
local isotropy group. As a result, vectors, tensors, etc. 
transform as
specific nontrivial representations of $U(2)$.

Consider the group $g \in SU(3)$ defined in its
fundamental representation as a
$3\times 3$ unitary matrix $g$ 
 of unit determinant. It
may be taken to be of the form $g = \exp ({i t_a \, \vf^a})$, where 
$t_a$ form a basis for traceless hermitian 
$3\times 3$ matrices, with $\Tr\,( t_a t_b) = \half \, \delta_{ab}$, and $\vf^a$ are the coordinates
for $SU(3)$.
The $SU(2)\subset U(2)$ subgroup is the standard
isospin subgroup, defined by the upper left
$2\times 2$ block, and corresponding to the generators
$t_a$, $a= 1, 2, 3$.
The $U(1)$ part of $U(2)$ is defined by the
hypercharge transformations,
with the generator $Y = 2\,t_8/\sqrt{3}$.
We also define right translation operators on $g$ by
$R_a\, g = g \, t_a$.
In terms of the frame fields $E^a_i$ for $SU(3)$, we may write these as
differential operators,
\beq
g^{-1} \, dg = - i t_a \, E^a_i \, d\vf^i,
\hskip .2in R_a = i (E^{-1})^i_a \, {\del \over \del \vf^i}, \hskip .2in
R_a \, g = g\, t_a
\label{cpII-2}
\eeq
Translations on $\mathbb{CP}^2$ correspond to the coset directions
$t_a$, $a= 4, 5, 6, 7$ and we can define the complex translation
operators as
\beq
R_{\pm 1} = R_4 \pm i R_5, \hskip .3in
R_{\pm 2} = R_6 \pm i R_7
\label{cpII-3}
\eeq
These will be
denoted by $R_i$, $R_{\bar i}$, $i, {\bar i} =
1, 2$. 

Functions on $\mathbb{CP}^2$ are invariant under
the $U(2)$ subgroup. So they admit a mode expansion of the form
\beq
F (g) = \sum_{s, A} C^{(s)}_{A} \, D^{(s)}_{A, w}(g)
= \sum_{s, A} C^{(s)}_{A} \, \la s, A \vert {\hat g} \vert s, w\ra
\label{cpII-4}
\eeq
Here
$\D_{AB}^{(s)}(g)= \la s, A \vert {\hat g} \vert s, B\ra$ are
the finite-dimensional unitary representation matrices for $SU(3)$
and in (\ref{cpII-4}) the 
states on the right, namely,
$\vert s, w\ra$ are invariant under
the $U(2)$ subgroup of $SU(3)$.
The action of $R_a$ on $\D_{AB}^{(s)}(g)$ is given by
\beq
R_a \,\la s, A \vert {\hat g} \vert s, B\ra
= \la s, A \vert {\hat g} \, T_a \vert s, B\ra
= \la s, A \vert {\hat g} \vert s, C\ra\, (T_a)_{CB}
\label{cpII-5}
\eeq
where $T_a$ are the matrix representatives of $t_a$ in the
representation designated by $s$. The invariance condition for functions
may therefore be stated as
\beq
T_a \, \ket{s, w} = 0, \hskip .2in a= 1, 2, 3, 8.
\label{cpII-6}
\eeq
In more detail, we can specify a state $\ket{s, A} = \vert a_1 a_2 \cdots a_p, b_1 b_2\cdots b_q\ra$ corresponding to a finite dimensional $SU(3)$ representation of the form
$T^{a_1 a_2 \cdots a_p}_{b_1 b_2 \cdots b_q}$, $a_i,\, b_j = 1,2, 3$.
We will refer to this as
a $(p,q)$-type representation. These are totally symmetric in all the upper indices
$a_i$'s and totally symmetric in all
the lower indices $b_j$'s with the trace (or any contraction between any choice of upper and lower indices) vanishing.
Under the action of $g\in SU(3)$,
$T^{a_1 a_2 \cdots a_p}_{b_1 b_2 \cdots b_q}$
transform as
\beq
T^{a_1 a_2 \cdots a_p}_{b_1 b_2 \cdots b_q}
\rightarrow  \bigl( g^{*a_1 a'_1}\,g^{*a_2 a'_2}\cdots \bigr)
\bigl( g_{b_1 b'_1}\,g_{b_2 b'_2}\cdots \bigr)
T^{a'_1 a'_2 \cdots a'_p}_{b'_1 b'_2 \cdots b'_q}
\label{cpII-7}
\eeq
For functions, we need $\la s, A \vert {\hat g} \vert s, w\ra$, with
states of the $(p,p)$-type, with $\ket{s,A}$ being general and 
\beq
\ket{s, w}  = \vert  3 3 3 \cdots,  3 3 3 \cdots \ra
\label{cpII-8}
\eeq

Vectors on $\mathbb{CP}^2$ must transform the same way as $R_{\pm i}$, so they must be doublets of $SU(2) \subset SU(3)$ and must carry hypercharge $Y = \pm 1$.
 Derivatives of functions of the form $R_{\pm i} f$ obviously satisfy this
 requirement.  Another possibility is given by representations of the type
 $(p+3, p)$ and $(p, p+3)$, so that a general parametrization of a vector 
 takes the form
 \beqar
A_i &=&- R_i \, f - \eta_{i {\bar i}} \,\epsilon^{{\bar i} {\bar j} } \sum_{s,A} C_A^{(s)} \,\la s, A\vert \,{\hat g}\, \vert  
{\bar j} 3 3 \cdots,  3 3 \cdots\ra\nonumber\\
\bA_{\bar i}  &=& -R_{\bar i} \, {\bar f} - \eta_{{\bar i}i} \,\epsilon^{i j} \sum_{s^*,A} C_A^{(s^*)} \,\la s^*, A\vert \,{\hat g} \,\vert 
3 3 \cdots, j 3 3 \cdots\ra
\label{cpII-9}
\eeqar
The particular state
$\vert  {\bar j} 3 3 \cdots,  3 3 \cdots\ra$ (of the $(p+3, p)$-type)
can be obtained, by the application of $R_{\bar j}$, from the
$SU(2)$ invariant states, with all indices equal to 3,
as
\beq
\eta_{i {\bar i}} \,\epsilon^{{\bar i} {\bar j} } \vert  {\bar j} 3 3 \cdots,  3 3 \cdots\ra
= \eta_{i {\bar i}} \,\epsilon^{{\bar i} {\bar j} } \,R_{\bar j} \,\vert
3 3 \cdots, 3 3\cdots\ra
\label{cpII-10}
\eeq
where $\eta_{i{\bar i}} = \delta_{i {\bar i}}$  is the metric
for $\mathbb{CP}^2$ in the tangent frame and
$\epsilon^{{\bar i} {\bar j} }$ is the Levi-Civita tensor.
With a similar result for the conjugate representation,
the parametrization (\ref{cpII-9}) can be written as
\beqar
A_i &=& -R_i \, f - \eta_{i{\bar i} } \epsilon^{{\bar i} {\bar j}} \, R_{\bar j} \chi
\nonumber\\
\bA_{\bar i} &=& -R_{\bar i} \, {\bar f}  - \eta_{{\bar i}i } \epsilon^{i j} \, R_{ j} {\bar\chi}
\label{cpII-11}
\eeqar
In a coordinate basis, rather than the tangent frames we have used above,
this becomes
\beq
A_k = -\nabla_k \,f + g_{k {\bar k}}  \bnabla_{\bar m}
\chi^{{\bar k} \bar m}, \hskip .3in
\bA_{\bar k}  = \bnabla_{\bar k}  \,{\bar f} - g_{{\bar k}k}  \nabla_{m}
{\bar \chi}^{{ k}  m}
\label{cpII-12}
\eeq
where $\chi^{{\bar k} \bar m}$ and ${\bar\chi}^{k m}$ are antisymmetric rank-two tensors, which for dimensional reasons, are proportional to
$\e^{{\bar k} \bar m}$ and $\e^{k m}$ and so can be reduced to
$\chi$ and ${\bar \chi}$. 

The nonabelian generalization of (\ref{cpII-12}) leads to the following
parametrization for the gauge potentials:
\beqar
A_i = - \nabla_i M \, M^{-1} + g_{i {\bar i}}  {\bar D}_{\bar j} \phi^{{\bar i} {\bar j}}\nonumber\\
\bA_{\bar i} = M^{\dagger -1} \bnabla_{\bar i} M^\dagger - g_{{\bar i} i}
D_j { \phi}^{\dagger ij}
\label{cpII-13}
\eeqar
In these expressions, $M$ and $M^\dagger$ are complex matrices which are elements of
$SL(N, \mathbb{C})$ if the gauge group is $SU(N)$; i.e., they are
$N \times N$ complex matrices with determinant equal to $1$.
(More generally, $M$, $M^\dagger$ will be in the complexification of the
gauge group.)
Further, $\phi^{{\bar i} {\bar j} } = \epsilon^{{\bar i} {\bar j} } \phi$,
$\phi^{\dagger i j} = \e^{i j} \phi^\dagger$ are also complex $N\times N$ matrices in the Lie algebra of $SL(N, \mathbb {C})$.
We also use an antihermitian basis for the fields, so that
$\bA_{\bar i} = - (A_i)^\dagger$.
The derivatives $D_j$ and $\bD_{\bar j}$ in (\ref{cpII-13})
are defined by
\beq
D_j \Phi = \nabla_j \Phi + [-\nabla_j M M^{-1}, \Phi],\hskip .3in
\bD_{\bar j} \Phi = \bnabla_{\bar j} \Phi + [
M^{\dagger -1} \bnabla_{\bar j} M^\dagger, \Phi]
\label{cpII-14}
\eeq
We have written these in terms of the action on
a field $\Phi$ (like $\phi$ or $\phi^\dagger$) which transforms under the adjoint representation
of the gauge group, i.e., as $\Phi \rightarrow U \, \Phi \, U^\dagger$,
where $U\in SU(N)$ is the gauge transformation.
The gauge transformation of the matrices $M$ and $M^\dagger$
is given by $M \rightarrow U\, M$, $M^\dagger \rightarrow M^\dagger U^\dagger$.
We can then see that 
the potentials in (\ref{cpII-13}) transform as connections.
It is sufficient to use just
$(- \nabla_j M \, M^{-1},  \, M^{\dagger -1} \bnabla_{\bar j} M^\dagger)$
in $D_j$ and $\bD_{\bar j}$  to ensure
that $D_j \Phi$ and $\bD_{\bar j} \Phi$ transform
covariantly under gauge transformations.
$(\nabla_j, {\bnabla}_{\bar j})$ are also taken to be Levi-Civita covariant,
so that 
(\ref{cpII-13}) behave correctly under
gauge and coordinate transformations.

It is possible to write the parametrization
(\ref{cpII-13}) in terms of manifestly gauge-invariant fields
by using the identities
\beqar
{\bar D} _{\bar j} \phi^{{\bar i}{\bar j}} &=& \bnabla_{\bar j} \phi^{{\bar i} {\bar j}}
+ [ M^{\dagger -1} \bnabla_{\bar j} M^\dagger, \phi^{{\bar i} {\bar j}} ]\nonumber\\
&=& M \left[ \bnabla_{\bar j} (M^{-1} \phi M)^{{\bar i} {\bar j}}  + [ H^{-1}\bnabla_{\bar j} H,
(M^{-1} \phi M)^{{\bar i} {\bar j}} ] \right] M^{-1}\nonumber\\
&=& M \left( {\bar \D}_{\bar  j} (M^{-1} \phi M )^{{\bar i} {\bar j}} \right) M^{-1}
= M \left( {\bar \D}_{\bar j} \chi^{{\bar i} {\bar j}} \right) M^{-1}
\label{cpII-15}\\
D_j {\phi}^{\dagger ij} &=&M^{\dagger -1}  \left( \D_j \chi^{\dagger i j} \right) M^\dagger
\nonumber
\eeqar
Here $\chi^{{\bar i} {\bar j}} = \epsilon^{{\bar i}{\bar j}} (M^{-1} \phi M )$,
${\chi}^{\dagger ij} = \epsilon^{ij} (M^{\dagger} \phi^\dagger  M^{\dagger -1})$ 
and
$H = M^\dagger M$.
The derivatives ${\bar \D}_{\bar j}$, $\D_j$ are defined
using the connections 
$H^{-1} \bnabla_{\bar j} H$, $-\nabla_j H H^{-1}$; i.e.,
\beq
{\bar \D}_{\bar j} \Phi = \bnabla_{\bar j} \Phi + [ H^{-1} \bnabla_{\bar j} H, \Phi],\hskip .2in
{\D}_{j} \Phi = \nabla_{ j} \Phi + [ -\nabla_j H H^{-1}, \Phi]
\label{cpII-16}
\eeq
By virtue of (\ref{cpII-15}), the parametrization (\ref{cpII-13}) can be
written as
\beqar
A_i &=& - \nabla_i M M^{-1} + M \left( g_{i {\bar i}} {\bar \D}_{\bar j} \chi^{{\bar i} {\bar j}} \right) M^{-1}  \nonumber\\
\bA_{\bar i} &=& M^{\dagger -1} \bnabla_{\bar i} M^\dagger
+ M^{\dagger -1} \left( -g_{{\bar i} i} \D_j \chi^{\dagger i j} 
\right) M^\dagger
\label{cpII-17}
\eeqar
Another equivalent version is given by
\beqar
A_i &=& -\nabla_i M M^{-1} - M a_i M^{-1}\nonumber\\
\bA_{\bar i} &=& M^{\dagger -1} \bnabla_{\bar i} M^\dagger 
+ M^{\dagger -1} \ba_{\bar i} M^\dagger
\label{cpII-18}\\
a_i &=& - M^{-1} g_{i{\bar i}} \bD_{\bar j} \phi^{{\bar i} {\bar j}} M=
- g_{i {\bar i}}{\bar \D}_{\bar j} \chi^{{\bar i} {\bar j}} \nonumber\\
\ba_{\bar i} &=& - M^{\dagger} g_{{\bar i} i} D_j \phi^{\dagger i j} M^{\dagger -1} =   - g_{{\bar i} i } \D_j {\chi}^{\dagger ij}  = {a_i}^\dagger
\nonumber
\eeqar
Notice that
$a_i$, $\ba_{\bar i}$ obey the conditions
\beq
g^{{\bar k} i } {\bar \D}_{\bar k} a_i = - {\bar \D}_{\bar i} {\bar \D}_{\bar j} \chi^{{\bar i} {\bar j}} = 0, \hskip .2in
g^{k {\bar i}} \D_k {\ba_{\bar i}} 
= 0,
\label{cpII-19}
\eeq
so that, effectively, they have only one independent component each.

The gauge-invariant degrees of freedom are given by
$H = M^\dagger M$ and 
$\chi = M^{-1} \phi \, M$,
${\chi^\dagger} = M^\dagger \phi^\dagger \, M^{\dagger -1}$.
Equivalently, they may be taken as
$H = M^\dagger M$ and $a_i$, $\ba_{\bar i}$, where
there are the additional constraints
(\ref{cpII-19}).
Yet another choice, also equivalent to the above mentioned ones,
would be
$\chi' =  M^\dagger \phi M^{\dagger -1}$,
$\chi'^\dagger = M^{-1} \phi^\dagger M$ and $H = M^\dagger M$.
From the point of view of carrying out the functional integration,
these fields are the coordinates for the gauge-orbit space $\C$.
A polar decomposition of $M$ as
$M = U \rho$, where $\rho$ is hermitian allows us to factor out the
gauge degrees of freedom and define a volume measure
on $\C = \A/ \G_*$. This will be taken up in the next section.

An interesting feature which is worthy of comment is
the holomorphic ambiguity or redundancy of
the parametrization (\ref{cpII-13}) or
(\ref{cpII-18}).
Notice that $(M, a_i, M^\dagger, \ba_{\bar i})$ and
$(M{\bar V}(\bx ), {\bar V}^{-1}(\bx) a_i {\bar V}(\bx), V(x) M^\dagger , V(x) \ba_{\bar i} V^{-1}(x))$ lead to the same gauge potentials, where
$V(x)$ is an $SL(N, \mathbb{C})$-matrix
with matrix elements which are holomorphic functions and ${\bar V}(\bx )$ is a similar antiholomorphic
matrix. Thus there is a certain ambiguity in how the matrices
$M$ and $M^\dagger$ can be chosen if $(A_i, \bA_{\bar i})$
are given.
On $\mathbb{CP}^2$, there are no globally defined holomorphic or
antiholomorphic functions, except for a constant.
So there are no additional degrees of freedom associated with this.
However, as mentioned in \cite{KMN1}, this feature may be useful with
locally defined $V$ and ${\bar V}$ to express fields in a nonsingular way in
various coordinate patches.

We close this section with another comment on the uniqueness of the
parametrization of the fields. This will be important for the
metric and volume element which we consider in the next section.
We have argued, based on the group theoretic counting of 
functional degrees of freedom, that any vector
can be parametrized as in (\ref{cpII-11});
hence any gauge potential on $\mathbb{CP}^2$ can be written in terms of $M$, $M^\dagger$, $\chi$,
$\chi^\dagger$.
The construction of $A_i$, $\bA_{\bar i}$ from the data
$(M, M^\dagger , \chi, \chi^\dagger )$ is thus clear.
Conversely, we can ask whether we can construct
$(M, M^\dagger , \chi, \chi^\dagger )$ from
$(A_i, \bA_{\bar i})$. This can indeed be done, as explained in some detail
in \cite{KMN1}. The fields $(M, M^\dagger , \chi, \chi^\dagger )$ will be nonlocal functions of $A_i, \, \bA_{\bar i}$
and their derivatives, consistent with the fact that
there is no way to factor out
the gauge degrees of freedom and obtain (unconstrained) gauge-invariant degrees of freedom in a local way in terms of $A_i, \, \bA_{\bar i}$. 

\section{The metric and volume element for $\C$}

We now turn to the metric on the space of gauge potentials ($\A$) and the
reduction of the associated volume element to the gauge orbit space
$\A/\G_*$. The starting point is the standard Euclidean metric on the space of 
the fields $A$, given by
\beq
ds^2 = - \int d\mu \, g^{i {\bar i}}\, \Tr ( \delta A_i \,\delta\bA_{\bar i} ) 
\label{cpII-20}
\eeq
Here $d\mu$ denotes the volume element for $\mathbb{CP}^2$.
In terms of the parametrization of the fields
(\ref{cpII-13}), we find
\beqar
\delta A_i &=&- D_i \theta + g_{i{\bar i}} \e^{{\bar i}{\bar j}}\left(
\bD_{\bar j} \delta \phi + [ \bD_{\bar j}\theta^\dagger, \phi]\right) \nonumber\\
&=&- D_i \theta  + [ \theta^\dagger , M a_i M^{-1}] + g_{i{\bar i}} \e^{{\bar i} {\bar j}} \bD_{\bar j} \delta\phi'\nonumber\\
\delta \bA_{\bar i} &=& \bD_{\bar i} \theta^\dagger
- g_{{\bar i} i } \e^{i j} \left( D_{j} \delta\phi^\dagger - [ D_j\theta, \phi^\dagger ]\right)\nonumber\\
&=&  \bD_{\bar i} \theta^\dagger + [ \theta , M^{\dagger -1} \ba_{\bar i} M^\dagger ] - g_{{\bar i} i} \e^{i j} D_j \delta\phi'^\dagger
\label{cpII-21}
\eeqar
where $\theta = \delta M M^{-1}$,
$\theta^\dagger = M^{\dagger -1} \delta M^\dagger$,
$\delta \phi' =\delta \phi + [\theta^\dagger ,\phi ]$,
$\delta \phi'^\dagger = \delta \phi^\dagger - [\theta, \phi^\dagger ]$.
We have also used the definition of $a_i$, $\ba_{\bar i}$ from
(\ref{cpII-18}). 
Upon using (\ref{cpII-21}) in (\ref{cpII-20}) and carrying out some integrations by parts, we find
\beqar
ds^2 &=& \int d\mu \,\Tr \Bigl[ 
\theta^\dagger ( - \bD \cdot D ) \theta + \delta \phi' ( -\bD \cdot D ) \delta \phi'^\dagger + \theta^\dagger g^{i{\bar i}} [ M a_i M^{-1}, [M^{\dagger -1} \ba_{\bar i} M^\dagger , \theta ] ] \nonumber\\
&&\hskip .6in+ \theta\, g^{ i {\bar i}} [ M^{\dagger -1} \ba_{\bar i} M^\dagger, D_i\theta ]
-\theta^\dagger\, g^{i {\bar i}} [ M a_{i} M^{-1} , \bD_{\bar i} \theta ]\nonumber\\
&&\hskip .6in
+\theta\, \e^{{\bar i} {\bar j}} [ M^{\dagger -1} \ba_{\bar i} M^\dagger , \bD_{\bar j} \delta\phi' ]
- \theta^\dagger\, \e^{i j} [ M a_{i} M^{-1} , D_{ j} \delta\phi'^\dagger]\,\Bigr]
\label{cpII-22}
\eeqar
We can write this as a quadratic form 
\beq
ds^2 = {1\over 2} \int d\mu~ \xi^{\dagger \alpha} \M_{\alpha \beta} \xi^\beta, \hskip .2in
\xi = (\theta , 
\theta^\dagger, \delta\phi', \delta\phi'^\dagger )
\label{cpII-23}
\eeq
We have written this in terms of components in the Lie algebra of
the gauge group $SU(N)$ by
 taking $\xi = \xi^\alpha (-i {t}^\alpha)$
where ${t}^\alpha$ form a basis for the Lie algebra.
It is useful to compare this with the metric
\beqar
ds^2_0 &=& {1\over 2} \int d\mu~ \xi^{\dagger\alpha}\, \xi^\alpha
=  \int d\mu~ \bigl[ \theta^{\dagger\alpha} \theta^\alpha + 
\delta\phi'^\alpha \delta \phi'^{\dagger\alpha} \bigr]\nonumber\\
&=& \int d\mu~ \left[ (M^{\dagger -1} \delta M^\dagger)^\alpha \,
(\delta M M^{-1})^\alpha + H^{\alpha \beta}_{\rm Adj}\,
\delta\chi'^\alpha \delta\chi'^{\dagger \beta} \right]
\label{cpII-24}
\eeqar
In the second line, we used the fact that $\delta \chi' = \delta (M^\dagger \phi M^{\dagger -1}) = M^\dagger \delta\phi' \, M^{\dagger -1}$ and its conjugate
version to rewrite the last term. Further $H^{\alpha\beta}_{\rm Adj}$ is the 
adjoint representation of $H$ defined by
$H^{\alpha \beta}_{\rm Adj} = - 2\, \Tr (H^{-1} {t}^\alpha H {t}^\beta )$.

From the structure of (\ref{cpII-23}), we see that the volume element 
takes the form $dV = \sqrt{\det \M} \, dV_0$, where $dV_0$ is the volume
element corresponding to (\ref{cpII-24}).
The first term in (\ref{cpII-24}) is the integral over $\mathbb{CP}^2$ of the Cartan-Killing metric for $SL(N, \mathbb{C})$. The corresponding volume element is thus the product over all points of the Haar measure
for $SL(N, \mathbb{C})$. Using the polar decomposition
$M = U \rho$, where $\rho$ is hermitian, and writing out the differential form of the top rank, we can see that this gives the volume element
$\prod_x [dU] d\mu(H)$, where $dU$ is the volume of $SU(N)$,
and $d\mu (H)$ is the Haar measure for $SL(N, \mathbb{C})/SU(N)$.
Further, since $H^{\alpha \beta}$ has unit determinant, we can write the
volume element of $ds^2_0$ as
\beq
dV_0 = \prod_x [dU]\, d\mu(H)\, d\chi' d\chi'^\dagger
\label{cpII-25}
\eeq
Finally going back to (\ref{cpII-23}), we get the corresponding volume element as
\beq
dV = \sqrt{\det \M} \,  \prod_x [dU]\, d\mu(H)\, d\chi' d\chi'^\dagger
\label{cpII-26}
\eeq
We can now factor out the volume of gauge transformations, i.e.,
factor out $\prod_x [dU]$ to obtain the volume element for
$\C = \A /\G_*$ as
\beq
d\mu [\C] = \sqrt{\det \M} \,  \prod_x d\mu(H)\, d\chi' d\chi'^\dagger
\label{cpII-27}
\eeq
The differentials appearing in this expression are for gauge-invariant fields.

The next step is the calculation of the determinant of $\M$.
For this, it is simpler to write the determinant as a functional integral
over a set of auxiliary fields
$B$, ${\bar B}$, $C$, ${\bar C}$. 
Explicitly,
\beq
{1\over \sqrt{\det \M}} = \int [dB d{\bar B} dC d{\bar C}]
e^{-S_0 - S_1}\label{cpII-28}
\eeq
where
\beqar
S_0&=& \int d\mu \left[ {\bar C}^\alpha (- \bD \cdot D )^{\alpha \beta}
C^\beta + {B}^\alpha (- \bD \cdot D )^{\alpha \beta}
{\bar B}^\beta\right]
\label{cpII-29}\\
S_1&=& \int d\mu \Bigl[ {\bar C}^\alpha ( M a M^{-1} \cdot M^{\dagger -1} 
\ba M^\dagger )^{\alpha \beta} C^\beta\nonumber\\
&&\hskip .4in+ C^\alpha (M^{\dagger -1} \ba M^\dagger \cdot D )^{\alpha\beta}
C^\beta + {\bar C}^\alpha (- M a M^{-1} \cdot \bD )^{\alpha\beta}
{\bar C}^\beta  \nonumber\\
&&\hskip .4in + C^\alpha (- \e^{{\bar i} {\bar j}} M^{\dagger -1} \ba_{\bar i} M^\dagger
\bD_{\bar j} )^{\alpha\beta} {B}^\beta
+ {\bar C}^\alpha (\e^{i j} M a_{i} M^{-1}
D_{ j} )^{\alpha\beta} {\bar B}^\beta\Bigr]
\label{cpII-30}
\eeqar
Equation (\ref{cpII-28}) expresses the required determinant in the 
form of a functional
integral for a standard field theory.
We have taken $C$ and ${\bar C}$ to behave like
$\theta$, $\theta^\dagger$, so they are scalar fields.
$B$ and ${\bar B}$ correspond to
$\delta\phi'$ and $\delta \phi'^\dagger$, so they are fields
with $Y = 2$ and $-2$, respectively.
The terms in $S_1$ involve powers of $a$, $\ba$, while the
differential operators in $S_0$ only depend on $H$ via the covariant derivatives. Our strategy will be to evaluate the integral in
(\ref{cpII-28}) as a perturbation series by expanding 
$e^{-S_1}$ in powers of $S_1$. The integral over
$e^{-S_0}$ will define the lowest
order result. Writing $\Gamma = \log \sqrt{\det \M}$, we find
$\Gamma = \Gamma_0 + \Delta \Gamma$, with
\beqar
e^{\Gamma_0} &=&\Bigl[\det (- \bD\cdot D)_{Y =0} \Bigr]\, \Bigl[\det (-\bD\cdot D)_{Y = -2}\Bigr]
\nonumber\\
e^{- \Delta \Gamma} &=& \bigl\langle e^{- S_1} \big\rangle
= \left[ {1 \over \int e^{-S_0}}\right] \int [dB d{\bar B} dC d{\bar C}]
e^{-S_0 - S_1}
\label{cpII-31}
\eeqar

The evaluation of the determinant $\det (- \bD \cdot D)_{Y = 0}$ was considered in \cite{KMN1}. This was done by writing its variation
as
\beq
\delta ( \Tr \log (- \bD\cdot D)_{Y=0} ) =
\int d\mu\, \Tr \left[ \delta (M^{\dagger -1} \bnabla M^\dagger ) M^{\dagger -1} \la 
{\hat J} \ra M^\dagger  + {\rm hermitian ~conjugate}\right]
\label{cpII-32}
\eeq
where
\beq
\la {\hat J} (x) \ra = - \D_x \G(x, y) \big\vert_{y \rightarrow x} ,
\hskip .3in
\G(x, y) = \left( {1\over -\bnabla\cdot \D}\right)_{x, y}
\label{cpII-33}
\eeq
The free ``propagator"
$G = (-\bnabla\cdot \nabla)^{-1}$ for scalar fields
on $\mathbb{CP}^2$ was calculated exactly.
 The propagator $\G(x, y)$ could then be obtained by
expansion in powers of $\nabla H H^{-1}$, since $\D$, as defined in (\ref{cpII-16}), involves this connection. 
  With regularization to take care of singularities as $y \rightarrow x$ in
the expression for $\la {\hat J} (x) \ra$, we obtained the leading terms
in the expansion of $\log\det (- \bD \cdot D)_{Y = 0}$ in 
terms of the monomials of $H$ and its derivatives of increasing scaling dimension.
The lowest order term was a WZW action for
$H$, somewhat surprisingly, with a finite coefficient.
The next set of terms of dimension four
were the $H$-dependent part of $\Tr (F^2)$ and $\Tr( g^{i{\bar i}} F_{i{\bar i}})^2$, with a logarithmically divergent coefficient.
There are also terms of higher dimension which will have finite coefficients, these were not explicitly calculated in \cite{KMN1}. We do not display the leading terms of $\det (- \bD \cdot D)_{Y =0}$ at this point, they will be given later, together with some of the other terms in (\ref{cpII-31}) which will be calculated presently.

The calculation of the leading terms in $\det (- \bD \cdot D)_{Y = -2}$
will proceed as in the case of $\det (- \bD \cdot D)_{Y = 0}$.
Since we will again expand in powers of $\nabla H H^{-1}$,
the key ingredient for this will be the free propagator
$(- \bnabla \cdot \nabla )^{-1}$, now for $Y = -2 $ fields.
This, as well as the issue of regularization for such propagators,
 will be taken up in the next section.
 
 Once we have the propagators, the calculation of terms resulting from
 $S_1$ will be straightforward. We will focus on the terms of scaling dimension $\leq 4$, corresponding to possible quadratic and logarithmic divergences. For these terms, it will suffice to consider terms up to four powers of $S_1$.
 
 \section{The propagator for $Y = -2$ fields}
 
 In this section, we calculate the propagator for fields with $Y = -2$, such as the field ${\bar B}$ in (\ref{cpII-29}), (\ref{cpII-30}).
 This field was identified as the variation of $\phi^\dagger$ in
 $\phi^{\dagger i j} =  \e^{ij} \phi^\dagger$, where $\phi^{\dagger i j}$ is a second rank antisymmetric tensor. The Laplace operator 
 on a second rank antisymmetric tensor with holomorphic indices
 will lead to the equation for the propagator for ${\bar B}$. Rather than using the tangent frame basis, if we use the coordinate basis for
 $\mathbb{CP}^2$, we can write
 \beq
 \delta \phi^{\dagger i j} = {\e^{i j} \over \sqrt{g}} \delta \phi^\dagger 
 = {\e^{i j} \over \sqrt{g}}  {\bar B}
 \label{cpII-34}
 \eeq
 Since $\mathbb{CP}^2$ is a K\"ahler manifold with potential
 $K = \log (1 + \bz \cdot z)$, the metric and
 Levi-Civita connections are easily worked out as
 \beqar
 g_{a\ba } &=& {\eta_{a \ba } \over (1+\bz\cdot z) } -{\eta_{a {\bar b} } \eta_{\ba b} \bz^{\bar b} z^b \over (1+ \bz\cdot z)^2}\nonumber\\
 g^{a \ba} &=& (1+ \bz\cdot z) \bigl( \eta^{a \ba} + z^a \bz^{\ba} \bigr)\nonumber\\
 \Gamma^{a}_{bc} &=& - {\left( \delta^a_b \eta_{c {\bar c}} +
 \delta^a_{c} \eta_{b {\bar c}} \right) \bz^{\bar c} \over (1+ \bz\cdot z) }, \hskip .2in
 \Gamma^{\ba}_{{\bar b} {\bar c} } =  (\Gamma^{a}_{bc} )^*
 \label{cpII-35}
 \eeqar
The other components of the connection vanish.
Using these results, we find
\beqar
\int d\mu \,g_{{\bar i} m} ( \bnabla_{\bar j} \delta \phi^{{\bar i} {\bar j}} )
( \nabla_n \delta \phi^{\dagger m n} ) &=&
\int d \mu \, B \left[ - (1+ \bz\cdot z) \bigl( \bdel\cdot \del + \bz\cdot \bdel \, z \cdot \del \bigr) + 3 + { 9 \over 4} \bz\cdot z\right.\nonumber\\
&&\hskip .6in\left. - {3\over 2} (1+ \bz\cdot z) \bigl( z\cdot \del -
\bz\cdot \bdel \bigr) \right] {\bar B}
\label{cpII-36}
\eeqar
This identifies the kinetic operator on ${\bar B}$.
The propagator thus obeys the equation
\begin{align}
\Bigl[ - (1+ \bz\cdot z) \bigl( \bdel\cdot \del + \bz\cdot \bdel \, z \cdot \del \bigr) + 3 + { 9 \over 4} \bz\cdot z &- {3\over 2} (1+ \bz\cdot z) \bigl( z\cdot \del -
\bz\cdot \bdel \bigr) \Bigr] \tilde{G}(z, y) \nonumber\\
&= {\delta^{(4)}(z, y) \over (\det g)}
\label{cpII-37}
\end{align}
where $\tilde{G}(z,y)$ is the free propagator for $Y=-2$ fields, to be distinguished from $G(z,y)$ which denotes the free propagator for scalars with $Y=0$.
In the case of scalar fields, we could take the propagator $G$ to be a function of the
distance between the two points corresponding to $z$ and $y$.
However, in the present case, there is an additional phase factor
due to the presence of the operator $z\cdot \del - \bz\cdot \bdel$
in (\ref{cpII-37}). This is ultimately due to $Y= -2$ for the field, which implies that it couples to the $Y$-component of the curvature and connection.
To isolate the phase factor, we note that the ${\bar B}$ field has the mode expansion
\beq
{\bar B} = \sum_{s, A} {\bar B}_A^{(s)} \, \bra{s, A} {\hat g} \ket{\underbrace{33 \cdots}_p;
\underbrace{33\cdots}_{p+3}}
\label{cpII-38}
\eeq
As a result, the propagator for the $Y=-2$ fields takes the form
\beqar
\tilde{G}(z, y) &=& \sum_p {d_p\over (p+2) (p+3)}
\bra{\underbrace{33 \cdots}_p;
\underbrace{33\cdots}_{p+3}} {\hat g}^\dagger_y {\hat g}_z \ket{\underbrace{33 \cdots}_p;
\underbrace{33\cdots}_{p+3}} \nonumber\\
&\sim&\left[ (g^\dagger_y g_z)_{33}\right]^3 \sum 
\left[ \left[(g^\dagger_y g_z)_{33} (g^\dagger_z g_y)_{33}\right]^p + \cdots\right]
\label{cpII-39}
\eeqar
where $d_p = \half (p+1) (p+4) ( 2 p +5)$ is the dimension of
the $(p, p+3)$-representation. In the second line of this equation, we have indicated the
result in terms of products of $g^\dagger_y g_z$ and its conjugate,
where $g$ is the $3\times 3$ matrix $g\in SU(3)$ used to
define local coordinates in the coordinate patch we are using.
There are also lower powers of $(g^\dagger_y g_z)_{33} (g^\dagger_z g_y)_{33}$ for each representation, as indicated by the
ellipsis in the square brackets. The point  is that there is
a common term $\left[ (g^\dagger_y g_z)_{33}\right]^3 $ which leads to
a phase factor. To see this more
explicitly, we note that $g \in SU(3)$ is
related to the local coordinates as
\beq
(g_z)_{i 3} = {z^i \over \sqrt{1 + \bz \cdot z}}, ~i = 1, 2,
\hskip .3in
(g_z)_{33} = {1 \over \sqrt{1 + \bz \cdot z}}
\label{cpII-40}
\eeq
The square of the distance, $s$, between two points with local coordinates $z^i$, $y^i$ 
can be taken to be
\beq
s (z, y) = \sigma^2_{z,y} = {1 \over (g^\dagger_y g_z)_{33} \, (g^\dagger_z
g_y)_{33}} - 1
= {(1+ \bz \cdot z) (1+ \by\cdot y) \over (1+ \by\cdot z) (1+ \bz\cdot y)}
- 1
\label{cpII-41}
\eeq
This is clearly translationally invariant (under left translations
$g \rightarrow u g $, $u \in SU(3)$) and agrees with
$s = \bz \cdot z$ when $ y = 0$, i.e., for $g_y = 1$.
Thus, we see from (\ref{cpII-39}) that $\tilde{G}$ is a function of
$s$, apart from the prefactor
\beqar
\left[(g^\dagger_y g_z)_{33}\right]^3 &=& \left[ (g^\dagger_y g_z)_{33}
\over (g^\dagger_z g_y)_{33}\right]^{3\over 2} \, \big\vert
(g^\dagger_y g_z)_{33}\big\vert^3
= \left[ (g^\dagger_y g_z)_{33}
\over (g^\dagger_z g_y)_{33}\right]^{3\over 2} \,{1\over (1+s)^{3\over2}}
\nonumber\\
&=& \left[ {(1+ \by \cdot z ) \over (1+ \bz \cdot y) }\right]^{3 \over 2}
\, {1\over (1+s)^{3\over 2}}
\label{cpII-42}
\eeqar
Combining this with (\ref{cpII-39}) we see that
a general ansatz for the propagator can be taken as
\beq
\tilde{G}(z, y) = \left[ {(1+ \by \cdot z ) \over (1+ \bz \cdot y) }\right]^{3 \over 2}\,
F(s)
\label{cpII-43}
\eeq
Having identified the phase factor, we can now use
(\ref{cpII-37}) to calculate $F(s)$.
For $z \neq y$, or $s\neq 0$, the $\delta$-function has no support and we can check that
the homogeneous equation is solved by
\beq
F(s) = C_1\, (1+ s)^{3\over 2} \left[ {1 + 2 s \over s (1+s)} + 2 \log \left( {s \over 1+s}\right) \right]
+ C_2\, ( 1+ s )^{3\over 2}
\label{cpII-44}
\eeq
where $C_1$, $C_2$ are arbitrary constants. Considering the short distance
behavior, we see that we need $C_1 = \half$ to reproduce the $\delta$-function on the right hand side of (\ref{cpII-37}).
Further, there should be no singularity as $s \rightarrow \infty$, this
identifies $C_2 = 0$. Combining (\ref{cpII-43}) and (\ref{cpII-44}), we get
the propagator for $Y = -2$ fields as
\beqar
\tilde{G}(z, y) &=& (1+ s)^{3 \over 2}  \left[ {1\over 2 s} +
{1  \over  2  (1+s)} +  \log \left( {s \over 1+s}\right) \right]\,
\left[ {(1+ \by \cdot z ) \over (1+ \bz \cdot y) }\right]^{3 \over 2}
\nonumber\\
&=& F(s)\, \left[ {(1+ \by \cdot z ) \over (1+ \bz \cdot y) }\right]^{3 \over 2}
\label{cpII-45}
\eeqar
where $s= \sigma^2_{z, y}$ is as given in
(\ref{cpII-41}). The last factor which is the phase can also be viewed as
arising from the path-ordered integral of the Levi-Civita connection
along a line joining $y$ and $z$.

\section{Regularization}

The regularization of the propagator for
$Y = -2$ fields will use the same point-splitting which was used for
scalar fields in \cite{KMN1}.
The key idea was to define the regularized version
of $G(z, y)$ as $G(z, y')$ where $y'$ specifies a 
point displaced from $y$ by a small distance of order $\sqrt{\e}$.
Thus $\e$ will serve as the regularization parameter, with $\e \rightarrow
0$ recovering the unregularized results. The shift
$y \rightarrow y'$ must be done in a way consistent with the isometries and gauge symmetries. 
It is useful to work this out in terms of the homogeneous coordinates
$Z = (Z_1, Z_2, Z_3)$, $Y=( Y_1, Y_2, Y_3)$ for the points
$z$, $y$, with the
identifications 
$Z \sim \lambda Z$, $Y \sim \lambda' Y$, $\lambda, \lambda' \in
\mathbb{C} - \{ 0\}$.
In terms of these coordinates, the distance between the points, given in (\ref{cpII-41}), can be
written as
\beq
s = \sigma^2_{z,y} = {{\bar Z} \cdot Z \, {\bar Y} \cdot Y \over {\bar Z} \cdot Y \, {\bar Y}\cdot Z} - 1
\equiv  \sigma^2 (Z,Y)
\label{cp2-38b}
\eeq
Notice that this expression has
invariance under the scaling $Z \rightarrow \lambda Z$,
$Y \rightarrow \lambda' Y$.
In a particular coordinate patch
with $Z^3, Y^3 \neq 0$, we can write
\beqar
Z &=& Z^3 (z^1, z^2, 1) = Z^3 \sqrt{1+ \bz \cdot z} ~ (g_{13}, g_{23}, g_{33})
\nonumber\\
Y &=& Y^3 (y^1, y^2, 1) = Y^3 \sqrt{1+ \by \cdot y} ~ (g'_{13}, g'_{23}, g'_{33})
\label{cp2-38c}
\eeqar
where $z^i = Z^i/Z^3$, $y^i = Y^i/Y^3$. 
We then recover the expression given in (\ref{cpII-41}).
With $s$ written in homogeneous coordinates as in
(\ref{cp2-38b}), (\ref{cpII-45}) gives a globally valid expression for the propagator.

For the
point-splitting regularization, we shift the point $Y$ to $Y'$, which is chosen to be
\beq
Y' = Y + \alpha \left( {W\, {\bar Y}\cdot Y \over {\bar Y}\cdot W}
- Y\right)
\label{cpII-46}
\eeq
Here $\alpha$ is a small complex number, $\vert \alpha \vert \sim \sqrt{\e}$, and $W$ parametrizes the
shift of coordinates.
The added term is chosen so as to have
the same scaling behavior as $Y$.
We then find  that
\beq
1 + \sigma^2 (Z, Y') =  {(1 + \sigma^2(Z,Y) ) (1 + \alpha {\bar\alpha} \sigma^2(Y, W)) 
\over \left[ 1 + \alpha \left( {{\bar Z}\cdot W \, {\bar Y}\cdot Y \over
{\bar Y}\cdot W  {\bar Z}\cdot Y} -1\right)\right]
\left[ 1 + {\bar\alpha} \left( {{\bar W}\cdot Z \, {\bar Y}\cdot Y \over
{\bar W}\cdot Y  {\bar Y}\cdot Z} -1\right)\right]}
\label{cpII-47}
\eeq
In  equation (\ref{cpII-45}), we can replace all factors
of $s$ by $\sigma^2(Z, Y')$ given by
(\ref{cpII-47}). Under coordinate transformations,
the propagator must transform as
$u^\dagger (z) \tilde{G}(z, y) u(y)$, where $u^\dagger$ is the hypercharge phase transformation with $Y = -2$. ($u$ will correspond to $Y = 2$.)
To preserve this property, we add an extra phase factor joining $y$ and $y'$.
We can view this extra term as a path-ordered integral for the Levi-Civita connection 
along a line joining $y$ and $y'$, analogous to the Wilson line for the gauge fields (see below)
that is needed to maintain gauge covariance
of the regulator.
Thus the UV regularized form of the propagator is given by
\beq
\tilde{G}_{\rm Reg}(z, y) = F( \sigma^2(Z,Y')) \,\left[ {(1+ \by' \cdot z ) \over (1+ \bz \cdot y') }\right]^{3 \over 2} \, \left[ {(1+ \bar{y} \cdot y' ) \over (1+ \by' \cdot y) }\right]^{3 \over 2}
\label{cpII-48}
\eeq
where as in \cite{KMN1} we will do a suitable angular averaging over the displacement due to point-splitting.
(It will turn out that, for some of the calculations we do, one of the
arguments of the propagator can be shifted to the origin by virtue of
translational invariance. The details of how this can be done
in terms of the homogeneous coordinates are given in
\cite{KMN1}; see also Appendix A of this paper.)

Since $\sigma^2 (Z, Y)$ and 
$\sigma^2 (Z, Y')$ are covariant quantities respecting the full isometry
(namely, left $SU(3)$ transformation on $Z$ or $Y$)
of $\mathbb{CP}^2$, the procedure we have outlined provides
a covariant point-splitting regularization.
However, we will need to modify this slightly to take account of 
covariance with respect to gauge transformations as well.
We have so far considered the
free propagator. In the presence of gauge fields, the propagator is $\tilde{\G}(z,y) = ((- \bD \cdot D )^{-1}_{z,y})_{Y=-2}$.
We carry out the explicit calculations by
expanding this in powers of the gauge field as
\beq
\tilde{\G}(z, y) = \tilde{G}(z, y) + \int_{y_1} \tilde{G}(z, y_1) {\mathbb V}_{y_1} \tilde{G}(y_1, y) +
\int_{y_1, y_2} \tilde{G}(z, y_1) {\mathbb V}_{y_1} \tilde{G}(y_1, y_2) {\mathbb V}_{y_2}
\tilde{G}(y_2, y) + \cdots
\label{cpII-53}
\eeq
where ${\mathbb V} = \bA\cdot \del + A \cdot \bdel + (\bdel\cdot A ) + \bA\cdot A$.
The propagator $\tilde{\G}(z, y)$ transforms as
$\tilde{\G} (z, y) \rightarrow U(z)\, \tilde {\G}(z, y)\, U^\dagger (y)$
under the gauge transformation $M \rightarrow U M$, $M^\dagger
\rightarrow M^\dagger U^\dagger$. This should be
respected in 
calculating currents such as $\la{\hat J} (z) \ra 
= - D_z \tilde{\G} (z, y)\bigr]_{y\rightarrow z}$ in (\ref{cpII-32}), (\ref{cpII-33}),
with regularization.
In other words, even though we shift $y$ to $y'$, the gauge transformation
must maintain the action of $U^\dagger (y)$ at the second argument
 to be consistent with its role in the current $\la{\hat J} (x) \ra $.
This means that a gauge-invariant point-splitting is given by
\beqar
\tilde{\G}_{\rm Reg} (z, y) &=& \tilde{\G} (z, y')\, \P \exp \left( -\int_y^{y'} ( M^{\dagger -1} \bnabla M^\dagger - \nabla M M^{-1} ) \right)
\nonumber\\
&=& \left[ \tilde{G}_{\rm Reg} (z,y) + \int_{y_1}\tilde{G}(z, y_1) {\mathbb V}(y_1) \tilde{G}_{\rm Reg}(y_1, y) 
+ \cdots\right]\nonumber\\
&&\hskip .2in \times \P \exp \left( -\int_y^{y+ \delta y} ( M^{\dagger -1} \bnabla M^\dagger - \nabla M M^{-1} ) \right)
\label{cpII-54}
\eeqar
Here $\tilde{G}_{\rm Reg}$ is as in (\ref{cpII-48}) and
$y' = y+\delta y$, with $\delta y^a \delta\by^{\bar a} \rightarrow \e\, \eta^{a {\bar a}}$ in taking the small $\e$ limit in a symmetric way.
(This is for the case when $y $ can be set to zero; which will cover
the calculations for which we need this factor.)
The path-ordered exponential helps to 
convert the $U^\dagger (y')$ (due to $\tilde{\G}(z, y')$) back to $U^\dagger (y)$.
(Such a step was already included for the isometries since
we added an extra phase factor in $\tilde{G}_{\rm Reg}$ connecting $y$ to $y'$.)
Because it involves the integral of
one-forms, it is adequate to use local coordinates $y$, $y'$ in (\ref{cpII-54}).

Turning to the infrared side of calculations, note that 
there are no infrared divergences since $\mathbb{CP}^2$ is a
compact space of finite volume.
However, we are only calculating the leading terms in $\Gamma$, up to
terms of scaling dimensions $\leq 4$. 
Terms of higher dimension will be ultraviolet finite. While they are irrelevant for issues of renormalization, we need to identify a kinematic
regime where such terms are suppressed to be able to make
any conclusions with the terms we calculate.
If we use an infrared cutoff $\lambda$, then the terms of higher scaling
dimension will carry inverse powers of this cutoff and will be suppressed for
modes of the fields with small momenta compared to $\lambda$.
This is what we do here. (This rationale for the infrared cutoff is explained in more detail in \cite{KMN1}.)
The WZW action will be special because it
is a term of lower dimension, {\it yet appears with a finite coefficient}.
Just for that particular term, we do the calculations both with and without
 an infrared cutoff. 

The details of the UV- and IR-regularized calculations will be given in the Appendix, but here we note that the IR regularization is easily
incorporated by using a simple integral representation for
 $F(s)$ in (\ref{cpII-45}) and including a lower cutoff. Explicitly, we write
\beqar
F(s) &=& {1 \over r^2} \, \int^{\infty}_{\lambda r^2} dt \, e^{-t s }
\left[ {1\over 2} \left(1+{s}\right)^{-{1\over2}} \right. \nonumber\\
&&\hskip .4in \left. + \left(1+{s }\right)^{{3\over2}} \left( {1 \over t} (e^{-t} -1) + e^{-t} \left(1 + {t \over 2} \right) \right) \right]
\label{cp2-53}
\eeqar
We have introduced $r^2$ via the scaling of coordinates. The infrared
cutoff $\lambda$ appears as the lower limit of the integration over $t$.
When $\lambda$ is set to zero, we clearly reproduce $F(s)$ in
(\ref{cpII-45}). This result, combined with (\ref{cpII-54}), can be
used for calculating the effective action.

\section {Results}

We are now ready to present the results regarding the volume element
for $\A/\G_*$. As mentioned at the end of section 3, we will consider the expansion of $\Gamma$ as a series in terms of increasing scaling dimension, focusing on those with dimension $\leq 4$. 
These are the terms we can expect to be relevant for the long wavelength modes of the fields; they are potentially UV-divergent terms, up to a possible logarithmic divergence.
For scaling dimension $2$, the possible terms correspond to a mass-like term for $a$, ${\bar a}$ (with a coefficient of order $1/\e$) and
a WZW term $S_{\rm wzw}(H)$, which, somewhat surprisingly,
has a finite coefficient.
There are also terms of dimension $4$ which arise with a coeffcient
of order $\log\e$.

\subsection {The WZW action}

All purely $H$-dependent terms, such as the WZW action, come from 
$\Gamma_0$ as defined in (\ref{cpII-31}). As mentioned above, the determinant
$\det(-\bar{D} \cdot D)_{Y=0}$ for scalar fields was already found in \cite{KMN1}. 
Using the result from there
\beqar
\Tr \log(-\bar{D} \cdot D)_{Y=0} &=& C_{Y=0} \, S_{\rm wzw} (H) ~+ \cdots 
\label{wzw1}\\
C_{Y= 0}&=&  {1\over \pi r^2} \left[ 1 - \log 2 + {3\over 2} e^{-\lambda r^2}
+ {\lambda r^2 \over 4} \right]\nonumber\\
&&\hskip .1in +{1\over \pi r^2}\left[ \left( E_1 (\lambda r^2) - E_1 (2 \lambda r^2)\right)
- {1\over 2} e^{-\lambda r^2} \left( 1- e^{-\lambda r^2}\right)\right]\nonumber\\
&&\hskip .1in +{1\over \pi r^2} \left[ {( 1- e^{-\lambda r^2})^2 \over 4 \lambda r^2}
+ \lambda r^2 ( e^{\lambda r^2} -1) E_1 (2\lambda r^2)
\right] 
\label{wzw2}
\eeqar
Here $S_{\rm wzw}(H)$ is the
four-dimensional
WZW action given by \cite{{Don},{NS}}
\beqar
S_{\rm wzw} (H) &=& {1\over 2 \pi} \int {\pi^2 \over 2} d\mu ~ g^{a \ba}\Tr ( \nabla_a H \, \bnabla_{\bar a} H^{-1} )
- {i \over 24 \pi} \int \omega\wedge \Tr (H^{-1} d H )^3 \nonumber\\
&=&{\pi\over 4} \int d\mu ~ g^{a \ba}\Tr ( \nabla_a H \, \bnabla_{\bar a} H^{-1} )
- {i \over 24 \pi} \int \omega\wedge \Tr (H^{-1} d H )^3
\label{wzw2a}
\eeqar
where $\omega$ is the K\"ahler two-form on $\mathbb{CP}^2$.
This can be expressed in local coordinates as
\beqar
\omega &=&{i }\, g_{a \ba}\, dz^a \, d\bz^{\ba} 
\label{wzw2b}
\eeqar
with $g_{a\ba}$ given by the Fubini-Study metric
(\ref{cpII-1}).  The last term in (\ref{wzw2a}) is, as usual, over a five-manifold
which has $\mathbb{CP}^2$ as the boundary.
We have normalized the volume of $\mathbb{CP}^2$ to $1$, so
there is an extra factor
of $\pi^2/2$ in (\ref{wzw2a}) compared to the standard normalizations.
(Also, we use a slightly different convention for
the normalization of $\omega$, compared to \cite{nair}.)

It is straightforward to simplify the expression for $C_{Y=0}$ in
(\ref{wzw2}) to
show that in the absence of an infrared cutoff ($\lambda \rightarrow 0$)
we get a finite result with no infrared divergence
\beq
C_{Y= 0} = {5 \over 2 \pi r^2}, \hskip .2in \lambda \rightarrow 0
\label{wzw3}
\eeq
On the other hand, for $\lambda r^2 \gg 1$
\beq
C_{Y=0} = {\lambda \over 4 \pi} + \mathcal{O}(1), \hskip .2in \lambda r^2 \gg 1
\label{wzw4}
\eeq

Turning to the $Y=-2$ part of $\Gamma_0$, i.e., $\Tr \log (-\bar{D} \cdot D)_{Y=-2}$, we can proceed 
analogously as in \cite{KMN1} by defining and calculating the current $\la {\hat J}(H, 1)\ra$. 
As for the scalar case
\beqar
\delta(\Tr \log (- \bD\cdot D)_{Y=-2} ) &=&
\int d\mu\, \Tr \left[ \delta (M^{\dagger -1} \bnabla M^\dagger ) M^{\dagger -1} \la 
{\hat J} \ra M^\dagger  + {\rm h. c.}\right] \nonumber\\
\la{\hat J} \ra &=& - \D_x \tilde{\G}(x, y) \big\vert_{y \rightarrow x} 
\label{wzw5}
\eeqar
where the only difference  
is that $\tilde{\G} (x,y) = ( -\bnabla \cdot \D)^{-1}_{x,y}$ is now defined 
as an expansion in free propagators for $Y=-2$ fields on $\mathbb{CP}^2$
(with the covariant derivatives $\nabla$ including the appropriate
Levi-Civita connections).

Expanding $\la {\hat J}\ra$ in powers of $\nabla H H^{-1}$, and using the UV and IR regularized propagators
as defined in (\ref{cpII-54}) and (\ref{cp2-53}), we can in principle find $\la {\hat J} \ra$
as we did for the scalar case. However, as the result is much more complicated for $Y=-2$
fields, instead, here we present the results in the two relevant limits, one with no infrared cutoff ($\lambda \rightarrow 0$) 
and the other for $\lambda r^2 \gg 1$:
\beqar
\la {\hat J}_a \ra &=&-{\pi \over 2} \, C_{Y=-2} \, \nabla_a H H^{-1} +\cdots \nonumber\\
\Tr \log(-\bar{D} \cdot D)_{Y=-2} &=& C_{Y=-2} \, S_{\rm wzw} (H) ~+ \cdots 
\label{wzw6}\\
C_{Y=-2} &=& {1 \over \pi r^2}, \hskip .2in \lambda \rightarrow 0 
\label{wzw7} \\
C_{Y=-2} &=& {\lambda \over 4 \pi} + \mathcal{O}(1), \hskip .2in \lambda r^2 \gg 1 
\label{wzw8}
\eeqar
The details of this calculation are given in Appendix A. 
We may also note that the 
 transition from the variation and the current in (\ref{wzw5}), (\ref{wzw6}) to the integrated version $S_{\rm wzw}(H)$ relies on the 
four-dimensional version of the Polyakov-Wiegmann identity,
which gives
\beq
\delta S_{\rm wzw} (H) = - {\pi\over 2} \int d\mu\, g^{a \ba} \Tr \left[ \bnabla_{\bar a} (\delta M^\dagger M^{\dagger -1})
\nabla_a H H^{-1} \right]
\label{wzw8a}
\eeq

As for the scalar case, other than 
the WZW action, $\Gamma_0$ will also include $\log \e$  terms (see below), and
finite terms as well (if we include terms with scaling dimension $> 4$, involving more derivatives on $H$). We will not calculate 
the finite terms in this paper.

Combining the results for the $Y = 0$ and $Y=-2$ fields, we have
\beqar
\Gamma_0 &=& \Tr\log(-\bD \cdot D)_{Y=0} + \Tr \log (-\bD\cdot D)_{Y=-2} \nonumber\\
&=& C \, S_{\rm wzw} (H) + \cdots
\label{wzw9}
\eeqar
where $C = C_{Y=0} + C_{Y=-2}$. 
In the case of no infrared cutoff the WZW action has a finite coefficient
\beq 
C = {7 \over 2 \pi r^2}, \hskip .2in \lambda \rightarrow 0
\label{wzw10}
\eeq
For $\lambda r^2 \gg 1$,
\beq
C = {\lambda \over 2 \pi}, \hskip .2in \lambda r^2 \gg 1
\label{wzw11}
\eeq

\subsection {The mass term}

Now we turn to terms with quadratic divergence. In $\Gamma_0$ the only term 
that could have $1 / \e$-divergence is the WZW action. However,
as we have seen,
for this term all divergences cancel out, giving an overall finite coefficient.

Among the next set of terms in $\Delta \Gamma$ in (\ref{cpII-31}),
there are a couple of possible candidates for quadratic divergence.
However, the only term which survives is a mass term for $a$ and $\ba$. This is expected as the only quadratically
divergent term in $\Delta \Gamma$ that is invariant under the holomorphic 
transformation $M^{\dagger} \rightarrow V M^{\dagger}$ is a mass term $\Tr [\ba H a H^{-1}]$.

Looking back at (\ref{cpII-31}) and expanding the exponential in $S_1$
we can write
\beq
\Delta \Gamma = \la S_1 \ra - {1 \over 2!} \la S_1^2 \ra + {1 \over 3!} \la S_1^3 \ra \cdots 
\label{mass1}
\eeq
where the sum is over connected diagrams. $S_1$ is defined in (\ref{cpII-30}), and
\beqar
\la C (x)\bar{C} (y)\ra &=& \left( {1\over - \bD \cdot D} \right)_{Y=0} = \G(x,y) \nonumber\\
\la \bar{B} (x) B (y)\ra &=&  \left( {1\over - \bD \cdot D} \right)_{Y=-2} = \tilde{\G} (x,y)
\label{mass2}
\eeqar

Since each term in the expansion of $\Delta \Gamma$ is a trace over the operators 
$M a M^{-1}$, $M^{\dagger -1} \ba M^\dagger$, $D$, $\bD$ and $(-\bD \cdot D)^{-1}$
in each such trace we can factor out $M^\dagger$, effectively setting $M^{\dagger} \rightarrow 1$
and $M \rightarrow H$.\footnote{Alternatively, in the path integral formulation of $e^{-\Delta \Gamma}$ 
one can make the transformations $C \rightarrow M^{\dagger} C$, $ \bar{C} \rightarrow \bar{C} M^{\dagger -1}$,
$B \rightarrow B M^{\dagger -1}$ and $ \bar{B} \rightarrow M^{\dagger} \bar{B}$. Since $\det M^{\dagger} = 1$ 
this redefinition of the fields does not affect the volume element
for the path-integral.} Quadratically divergent terms will come only from the first two terms 
in the expansion.
Performing similar calculations as in the case of the current in $\delta \Gamma_0$, we get
\beq
\Delta \Gamma = \left(-{1 \over 4 \e} + {1 \over 2 r^2} \log \e \right) \int d \mu  \, g^{a \ba} \Tr ( \ba_{\ba} H a_a H^{-1} ) +\mathcal{O}(\log \e)
\label{mass3}
\eeq
The only quadratically divergent term is thus a mass term for $a$ and $\ba$. 

\subsection {The log-divergent terms}

Logarithmically divergent terms can arise from both $\Gamma_0$ and $\Delta \Gamma$. Combining all such contributions,
\beqar
\Gamma_{\log \e} &=& \Gamma_{0 \, \log \e} + \Delta \Gamma_{\log \e} \nonumber\\
&=& {\log \e \over 12} \int \Tr \, \Bigl[ (g^{a \ba} \bnabla_{\ba} (\nabla_a H H^{-1} ) )^2
 +(g^{a \ba} \ba_{\ba} H a_a H^{-1} )^2  \nonumber\\
&& \hskip .8in  + g^{a\ba} g^{b\bar{b}}[\ba_{\ba}, H a_a H^{-1}] \bnabla_{\bar{b}}(\nabla_b H H^{-1}) \nonumber\\
&& \hskip .8in   -g^{a\ba} g^{b\bar{b}} \left( \bnabla_{\ba} ( \nabla_b H H^{-1} ) [\ba_{\bar{b}} , H a_a H^{-1} ]
+ \bnabla_{\ba} \ba_{\bar{b}} \D_a (H a_b H^{-1}) \right) \Bigr]
\label{log1}
\eeqar
where the trace is in the adjoint representation. (Once again, the details
are given in Appendix B.)

\section {Discussion}

The key results we have obtained are the following.
We introduced a parametrization of the gauge potentials,
see equation (\ref{cpII-18}) in section 2, which allows for the explicit
factoring out of gauge transformations and consequent reduction to
gauge-invariant degrees of freedom.
In particular, the volume element for the gauge orbit space
$\C = \A /\G_*$ can be written as
\beq
d\mu [\C] = e^\Gamma \,  \prod_x d\mu(H)\, d\chi' d\chi'^\dagger
\label{cpII-74}
\eeq
Unlike the case of two dimensions, an exact calculation of $\Gamma$ is not
possible in four dimensions. We have calculated the first set of terms
in $\Gamma$ corresponding to monomials of the fields and their derivatives
of scaling dimension $\leq 4$. These are the terms which can be expected to be
potentially ultraviolet divergent.
The first of these terms is a four-dimensional WZW action
$S_{\rm wzw}(H)$ for
 $ H \in SL(N, {\mathbb C})/ SU(N)$ (for an $SU(N)$ gauge theory),
given in (\ref{wzw9}). 
{\it A priori}, on dimensional grounds, one may expect this term to
be quadratically divergent, but, somewhat surprisingly, it arises with
a finite coefficient.
The sign of the coefficient is ``correct" in the sense of ensuring
convergence for functional integration
over $H$. (In \cite{KMN1}, we calculated a similar contribution due to
chiral scalar fields on $\mathbb{CP}^2$. The 
coefficient of $S_{\rm wzw}(H)$ was such that it tends to destabilize the
theory for long wavelength modes of
$H$. It is interesting that chiral scalars on $\mathbb{CP}^2$ have
this destabilizing effect. For the volume element for
$\C$, there is no such issue.)

The second term in $\Gamma$ is a mass term for 
the components $a_i$, ${\ba}_{\bar i}$ of the potentials, which are related to
$\chi'$, $\chi'^\dagger$ as
$a_i = - g_{i {\bar i}} \e^{{\bar i} {\bar j}} H^{-1} ({\bar \nabla}_{\bar j} \chi' ) H
$, 
${\ba}_{\bar i} = - g_{ {\bar i} i} \e^{i j}  H (\nabla_{j} \chi'^\dagger) H^{-1} $.
The coefficient of this term, say $\mu^2_{\rm div}$,
is quadratically divergent. 
Since ultraviolet divergences are related to products of operators at the same
point, i.e., to ultralocal geometry and hence not sensitive to global geometry, the divergence shows that
this term should survive even if we take the large $r$ limit of
$\mathbb{CP}^2$.
Further, since it is consistent with gauge invariance and all the isometries of $\mathbb{CP}^2$, there is no
reason to reject this. This means that the functional measure
has to be defined by introducing a similar counterterm 
(with a coefficient $\mu^2_{\rm counter}$) from the beginning
and then renormalizing by setting $\mu^2_{\rm div} + \mu^2_{\rm counter}$
to a finite value $\mu^2_{\rm Ren}$.
This finite renormalized value $\mu_{\rm Ren}$ has the dimensions
of mass and will serve as the mass parameter defining the theory.
This is in accordance with the fact that dimensional transmutation,
with the introduction of an arbitrary scale factor,
is needed to define four-dimensional gauge theories.
In the usual perturbative approach, this would enter via
loop corrections and the running coupling constant, but in
our formulation, it appears as what is 
needed to make the volume 
element for $\C$ (or the measure of functional integration)
well-defined.
If loop calculations are carried out in our approach, 
the results will be functions of $\mu_{\rm Ren}$; there will be no need for
additional dimensional parameters; what is conventionally
considered as $\Lambda_{\rm QCD}$ will be related to
$\mu_{\rm Ren}$.

There is also a set of terms which are of scaling dimension
$4$ with a logarithmically divergent coefficient, see
(\ref{log1}).
Unlike the case of flat space, these terms do not combine into
$\Tr (F^2)$ since the reduced isometries of
$\mathbb{CP}^2$ allow
for additional tensor structures.
Presumably, to make the volume element well-defined,
similar counterterms have to be introduced {\it a priori}
and renormalization has to be carried out.
Finally there is also an infinity of terms of scaling dimension
$> 4$, which are ultraviolet finite, which we have not calculated.
They are presumably less relevant to the dynamics of long wavelength modes
of the fields compared to the terms we have calculated.

Returning to the mass term (\ref{mass3}), 
we note that the possibility of a soft gluon mass has been proposed already
in the
1980s \cite{cornwall}.
Lattice simulations of the gluon propagator
in the Landau gauge also indicate
its saturation to a finite value at low momenta, consistent with
a propagator mass  \cite{gmass-lat}.
At least qualitatively, we need an analytic
understanding of these
lattice results.
 Indeed, a number of papers have analyzed the Schwinger-Dyson equations of QCD
 with a view to showing that the gluon self-energy is nonvanishing at
 zero momentum, along with attempts to extract quantitative
 predictions from it \cite{{CPB},{ABP}}.
The appearance of a possible gauge-invariant
mass term in our analysis provides a parallel track of viewing such
analyses.

Perhaps the most striking and qualitatively new feature of our analysis is
the appearance of the WZW action $S_{\rm wzw}(H)$. With the above given argument
for a nonzero mass term (\ref{mass3}), it is then possible to consider
a kinematic regime of momenta $\ll \mu_{\rm Ren}$
where we can neglect the massive components $a_i$, $\ba_{\bar i}$ and
consider a reduced theory where
\beq
A_a \simeq - \nabla_a M \, M^{-1}, \hskip .2in
\bA_{\bar a} \simeq M^{\dagger -1} {\bar \nabla}_{\bar a} M^\dagger
\label{disc1}
\eeq
The volume element then takes the form
\beqar
d\mu [ \C ] &=& e^\Gamma\,  d\mu (H) \nonumber\\
\Gamma&\simeq& C  \, S_{\rm wzw}(H) + C_1 \int \Tr \left(g^{{\bar a} a}  \bnabla_{\bar a}
(\nabla_a H H^{-1}) \right)^2 + \cdots \nonumber\\
&\approx& C  \, S_{\rm wzw}(H) + \cdots
\label{disc2}
\eeqar
(Here $C_1$ is the renormalized value of the coefficient of the term
in (\ref{log1}), after the $\log \e$-divergence is eliminated.)
In the last line of (\ref{disc2}) we have neglected the term quartic in the derivatives as it is less significant for long wavelength modes 
compared to $S_{\rm wzw}(H)$.
The theory defined by (\ref{disc2}) should be applicable for $\lambda \ll \mu^2_{\rm Ren}$,
with $\lambda r^2 \gg 1$.
This theory is the four-dimensional WZW theory on
$\mathbb{CP}^2$. So our conclusion is that we expect that
for fields of modes of wavelength small compared to
$\mu_{\rm Ren}$, the four dimensional Yang-Mills theory can be
approximated by a 4d-WZW theory for the field
$H \in G^{\mathbb{C}}/G = SL(N, \mathbb{C}) / SU(N)$.

The 4d-WZW theory, we may note, also has
a history going back to the 1980s, appearing first
in the work of
Donaldson in the context of antiself-dual
instantons \cite{Don}.
The same theory is obtained
in the K\"ahler-Chern-Simons theory \cite{NS} which
attempted
to generalize the 2d-WZW theory to four dimensions,
a paradigm similar to the
WZW-CS relation in two and three dimensions \cite{witten}. 
As shown in \cite{NS}, and elaborated in \cite{{Los}, {ueno}},
this action also leads to
a holomorphically factorized current algebra, analogous to the
case
in two dimensions. 4d-WZW theories have also been found in higher dimensional quantum Hall systems
\cite{KNqhe}.
They also describe the target space dynamics of (world-sheet) ${\cal N} =2$ heterotic superstrings
\cite{OV}. More recently, such theories have been analyzed
 in \cite{costello}
in the context of holomorphic field theories on
twistor space.

The critical points of the action $S_{\rm wzw}(H)$
are antiself-dual instantons. They are related to
holomorphic vector bundles, 
with $M$ and
$M^\dagger$ defining the holomorphic frames for the bundle.
What is interesting is that there is some evidence, based on
lattice simulations, that
the correlation functions for
gauge fields and hadrons seem to be dominated by instantons
at low energies; see, for example,
\cite{{schafer},{athenodorou}}.
While it is difficult to see instanton dominance
analytically for fields on
$\mathbb{R}^4$, the present result that
the theory can be approximated by the 4d-WZW theory
along the lines argued above provides
some 
analytical evidence for an instanton liquid picture.

Finally, there is another aspect of the 4d-WZW theory
which is worth pointing out.
In the (2+1)-dimensional analysis considered in
\cite{{KKN},{nair-trento1}}
the expectation value of the Wilson loop operator (in a representation indicated as $R$) takes the
form
\beqar
\la W_R(C)\ra &=& {\cal N} \int d\mu (H) e^{ 2 c_A S^{(2d)}_{\rm wzw} (H)}
\, \exp\left( - {8\pi \over e^4 c_A}\int \Tr \bigl(\bnabla (\nabla H H^{-1})\bigr)^2
\right) \, \Tr\left[\P e^{ \oint_C \nabla H H^{-1} }\right]\nonumber\\
&\sim& e^{- \sigma_R {\rm Area}(C)}, \hskip .3in \sigma_R = e^4 {c_A c_R \over 4 \pi}
\label{disc3}
\eeqar
where $S^{(2d)}_{\rm wzw} (H)$ is the 2d-WZW action for $H$,
$c_R$, $c_A$ are the values of the quadratic Casimir operators for
the representation $R$ and for the adjoint representation,
respectively. $e^2$ is the coupling constant of the Yang-Mills theory.
Notice that, as $e^2 \rightarrow \infty$, which is
the limit where the integrand in (\ref{disc3}) defines the
2d-WZW theory for $H$,
the expectation value
of $W(C)$ vanishes for any curve $C$ enclosing any nonzero area.
If we consider evaluating $\la W(C)\ra$ in terms of correlators
for the current $\nabla H H^{-1}$,
the leading term due to the two-point function is of the form
\beq
\oint\oint dz dz' \la \nabla H H^{-1}(z) \, \nabla H H^{-1} (z')\ra
\sim - c_R \oint\oint {dz dz' \over (z- z')^2}
\label{disc4}
\eeq
The UV singularity of this integral is not regularized when
$e^2 \rightarrow \infty$, and this is the genesis of the
vanishing of $\la W(C)\ra$.

We see that a similar situation is obtained in the theory
defined
in the 4d-theory (\ref{disc2}). Using the Polyakov-Wiegmann identity
\beq
S_{\rm wzw}(NH) =  S_{\rm wzw}(H)  + S_{\rm wzw}(N) 
- {\pi \over 2} \int g^{{\bar a} a}  \Tr ( N^{-1} \bnabla_{\bar a} N \,
\nabla_a H H^{-1} )
\label{disc5}
\eeq
we obtain
\beq
\int d\mu (H) \, \exp\left(
C S_{\rm wzw}(H) - C {\pi \over 2} \int g^{{\bar a} a}  \Tr ( N^{-1} \bnabla_{\bar a} N \,
\nabla_a H H^{-1} ) \right) 
= e^{ - C S_{\rm wzw} (N)}
\label{disc6}
\eeq
By taking small variations of $N$, we then find
\beq
\left< g^{{\bar a}  a} \bnabla_{\bar a}  (\nabla_a H H^{-1})^\alpha (x)
g^{{\bar b} b} \bnabla_{\bar b}  (\nabla_b H H^{-1})^\beta (y)
\right> = - {4\over \pi C} g^{{\bar a} a} \bnabla_{{\bar a}, y} \nabla_{a, y}
\delta (y, x)\, \delta^{\alpha \beta}
\label{disc6}
\eeq
The two-point function for the currents can be obtained from this
as
\beq
\left<  (\nabla_a H H^{-1})^\alpha (x)  (\nabla_b H H^{-1})^\beta (y)
\right> = {4  \over \pi C} \, \nabla_{a, x} \nabla_{b, y} G(y, x) \delta^{\alpha\beta}
\label{disc7}
\eeq
where $G(y, x)$ is the propagator for scalars on $\mathbb{CP}^2$
given in \cite{KMN1} as
\beq
G(y,x) = {1\over 2 \, s}   - {1\over 2} \log\left( {s \over 1+s}\right)
- {3\over 4}, 
\label{disc8}
\eeq
where $s = \sigma^2_{y,x}$ is given by (\ref{cpII-41}).
We see that 
the logarithmic term of $G(y, x)$ in (\ref{disc7}) can 
indeed reproduce a result similar
to what was obtained in (2+1) dimensions.
It is not possible to make a more complete analysis at this stage,
but clearly the similarity with
(2+1) dimensions shows that
the reduced theory (\ref{disc2}) in terms of the
4d-WZW action is worthy of further investigation. This will be
left to future work.

\bigskip

This work was supported in part by the U.S. National Science Foundation Grants No. PHY-2112729 and No. PHY-1915053, PSC-CUNY grants and by
the Bernard B. Levine Graduate Fellowship at the City College of
New York.

\section*{Appendix A: Calculating $ \la {\hat J} \ra$ for the $Y=-2$ fields}
\def\theequation{A\arabic{equation}}
\setcounter{equation}{0}
In this Appendix we outline the calculations of the current $\la {\hat J} \ra$
in (\ref{wzw5}) for $Y=-2$ fields. The procedure is analogous to the scalar 
calculation in \cite{KMN1}. The only difference is in the propagator
and the connections of the gradients as given in (\ref{cpII-35}).

The current is
\beqar
\la {\hat J} \ra &=& - \D_x \tilde{\G}_{\rm Reg} (x,y) \big\vert_{y\rightarrow x} \nonumber\\
&=&{\rm Term~1} + {\rm Term~2} + {\rm Term~3} + \cdots
\nonumber\\
{\rm Term~1}&=& - \nabla_{xa} \tilde{G}(x,y')\, \left[ {(1+y' \cdot \bar{y}) \over (1+\bar{y}'\cdot y) }\right]^{{3 \over 2}} \,
 \P \exp\left( \int_y^{y'} \nabla H H^{-1} \right)\big\vert_{y\rightarrow x}\nonumber\\
 {\rm Term~2}&=& (\nabla_a H H^{-1})_{x} \, \tilde{G} (x, x') \, \left[ {(1+x' \cdot \bar{x}) \over (1+\bar{x}'\cdot x) }\right]^{{3 \over 2}}
 \label{A1}\\
 {\rm Term~3}&=& \int_z \nabla_{xa} \tilde{G}(x, z)g_{z}^{b {\bar b}} (\nabla_b H H^{-1} )_{z} \bnabla_{z \bar b} \tilde{G} (z, x') \, \left[ {(1+x' \cdot \bar{x}) \over (1+\bar{x}'\cdot x) }\right]^{{3 \over 2}} \nonumber
 \eeqar
The primed coordinate is given in (\ref{cpII-46}) and we perform an angular integration over
$w$ and $\alpha$ with the conditions that $\vert\alpha\vert^2 = \epsilon$ and $\sigma^2(x,w) = 1$.

As in the scalar calculation (in \cite{KMN1}), we make a coordinate transformation
$w \rightarrow w'$ such that $w^a = x^a + (e^{-1}_x)^a_b { w'^b \over 1- \bar{x}\cdot w'}$,
where $e^{-1}$ are the tangent frame fields. Under this transformation $\sigma^2(x,w) = \vert w'\vert^2$.
${\rm Term~1}$ in (\ref{A1}) becomes
\beqar
{\rm Term~1} &=&  - \int_{\alpha} \delta (\vert\alpha\vert^2 - \e) \int_{w} \delta(\sigma^2(x,w) -1) \biggl\{ \nabla_{xa} \tilde{G}(x,y')\, \left[ {(1+y' \cdot \bar{y}) \over (1+\bar{y}'\cdot y) }\right]^{{3 \over 2}} \nonumber\\
&&\hskip .5in \times
(y'-y)^b (\nabla_b H H^{-1})_y \big\vert_{y \rightarrow x}\biggr\} \nonumber\\
&=& (\nabla_a H H^{-1})_x {\e \over 2} \left( (1+\e) F'(\e) - {3\over2} F(\e) \right)
\label{A2}
\eeqar
where the extra term $-{3\over2} F(\e)$ (as compared to the scalar case) 
comes from the spin connection in $\nabla_a$ and the phase factor of $\tilde{G}(x,y')$.
Including the scale $r$ in (\ref{A2}), we take $\e \rightarrow \e / r^2$ and
$F(s)$ is the IR regulated propagator
\beqar
F(s) &=& {1 \over r^2} \, \int^{\infty}_{\lambda r^2} dt \, e^{-t s }
\left[ {1\over 2} \left(1+{s}\right)^{-{1\over2}} \right. \nonumber\\
&&\hskip .4in \left. + \left(1+{s }\right)^{{3\over2}} \left( {1 \over t} (e^{-t} -1) + e^{-t} \left(1 + {t \over 2} \right) \right) \right]
\label{A3}
\eeqar

${\rm Term~1}$ then becomes
\beq
{\rm Term~1} =  (\nabla_a H H^{-1})_x \left(- {1 \over 4 \e} - {1 \over 8 r^2} \right)
\label{A4}
\eeq

Performing similar 
calculations as for ${\rm Term~1}$ it's straightforward to 
find that
\beqar
{\rm Term~2} &=& (\nabla_a H H^{-1})_x \left[{1 \over 2\e} + {1 \over r^2} \log \left( {\e \over r^2 } \right) - {\lambda \over 2}  - {1 \over 4 r^2} + {1 \over 2 r^2} e^{-\lambda r^2} (3+\lambda r^2)  \right. \nonumber\\
&& \hskip .9in \left.+{1 \over r^2} \left( E_{1}(\lambda r^2) + \log (\lambda r^2) + \gamma \right) \right]
\label{A5}
\eeqar
where $E_1$ is the exponential integral
\beq
E_1(w) = \int_1^{\infty} \, {dt \over t} e^{-wt}
\label{A6}
\eeq

For ${\rm Term~3}$ we need three coordinate transformations:\\
1) A transformation $w \rightarrow w'$ such that $\sigma^2(x,w)=1$, as explained above;\\
2)  A transformation $ z \rightarrow z'$ such that  $z^a = x^a + (e^{-1}_x)^a_b { z'^b \over 1- \bar{x}\cdot z'}$, setting $\sigma^2(x,z) = \vert z'\vert^2$;\\
3) Finally, a transformation $z' \rightarrow \tilde{z}$, which can be given in homogenous coordinates as
${\tilde{Z} / {\tilde Z}_3}= {\bar{\tilde W}}_3 {Z' / (\bar{\tilde{W'}}\cdot Z'})$, where $\tilde{W'} = (\alpha W'_1, \alpha W'_2, W'_3) = W'_3 (\alpha w'_1, \alpha w'_2, 1)$. \\
The first two transformations effectively eliminate $x$ from the integrals by
translating the integration variables. The last one is useful because $1 + \sigma^2(z',\alpha w') = (1+\e)(1+\vert\tilde{z}\vert^2)$
which significantly simplifies the integrals in $\tilde{z}$ and $w'$. These transformations are the same as in \cite{KMN1} (where more details are given as well) and, as was
the case there, the integration measures remain unchanged.
With these changes, ${\rm Term~3}$ becomes
\beqar
{\rm Term~3} &=&  (\nabla_b H H^{-1} )_{x} \int_{\a} \delta (\vert\a\vert^2 -\e) \int_w \delta( \sigma^2(x,w) -1) \nonumber\\
&& \hskip .4 in \int d\mu (z)  \,  \nabla_{xa} \tilde{G}(x, z)\, g_{z}^{b {\bar b}} \, \bnabla_{z \bar b} \tilde{G} (z, x') \, \left[ {(1+x' \cdot \bar{x}) \over (1+\bar{x}'\cdot x) }\right]^{{3 \over 2}}
\nonumber\\
&=& - (\nabla_b H H^{-1} )_{x}\nonumber\\
&&\times \int d\mu ({\tilde{z}}) \,  (1+ \vert\tilde{z}\vert^2) (e^{-1}_x)^b_m \tilde{z}^m \,\eta_{a \bar{a}} (e_x)^{\ba}_{\bar{m}}  \bar{\tilde{z}}^{\bar{m}}
\, \left(F'(\vert\tilde{z}\vert^2)(1+\vert\tilde{z}\vert^2) - {3 \over 2} F(\vert\tilde{z}\vert^2) \right)  \nonumber\\
&& \hskip 1.1 in \left( F'(\vert\tilde{z}\vert^2(1+\e) +\e) (1+\vert\tilde{z}\vert^2)(1+\e) + {3\over2} F(\vert\tilde{z}\vert^2(1+\e) +\e) \right) \nonumber\\
&=&- (\nabla_a H H^{-1} )_{x} \int^{\infty}_0 ds \, s^2 \, \left (F'(s) - {3 \over 2(1+s) } F(s) \right) \nonumber\\
&& \hskip 1.1in  \left(F'(s(1+\e) +\e)(1+\e) + {3 \over 2(1+s)} F(s(1+\e) +\e) \right) \nonumber\\
&=&  \mathcal{I} \, (\nabla_a H H^{-1} )_{x} 
\label{A7}
\eeqar
The exact calculation of $\mathcal{I}$ is more complicated than for 
its scalar counterpart. So instead of calculating it for arbitrary $\lambda$,
we find it in the two limits, namely, in the case of no infrared cutoff, i.e., 
$\lambda \rightarrow 0$,
and in the case of $ \lambda r^2 \gg 1$. Including the scale factor
$r$, we find
\beqar
\mathcal{I} &=& -{1 \over 4 \e} - {1 \over r^2} \log \left( {\e \over r^2} \right) -{13 \over 8 r^2}, \hskip .7in \lambda \rightarrow 0
\nonumber\\
\mathcal{I} &=& - {1 \over 4 \e} - {1 \over r^2} \log \left( \lambda \e \right) + {3 \lambda \over 8} + \mathcal{O}(1), \hskip .3in \lambda r^2 \gg 1
\label{A8}
\eeqar
Combining (\ref{A4}), (\ref{A5}), and (\ref{A8}), and taking the appropriate limits,
\beqar
\la {\hat J}_a \ra &=& - {\pi \over 2} \, C_{Y=-2} \, \nabla_a H H^{-1} + \cdots \nonumber\\
C_{Y=-2} &=& {1 \over \pi r^2}, \hskip .8 in \lambda \rightarrow 0 \nonumber\\
C_{Y=-2} &=& {\lambda \over 4 \pi} + \mathcal{O}(1), \hskip .3in \lambda r^2 \gg 1 
\label{A9}
\eeqar
which are the results in (\ref{wzw7}) and (\ref{wzw8}).

\section*{Appendix B: Ultraviolet divergent terms}
\def\theequation{B\arabic{equation}}
\setcounter{equation}{0}
In this appendix we go over some of the calculations 
leading to the mass term and log-divergent terms in (\ref{mass3}) and (\ref{log1}).
For the mass term we proceed as in Appendix A. 
As for terms that have at most a log-divergence we calculate them 
in the flat limit ($r \rightarrow \infty$), afterwards restoring
$r$ in the metric and the volume element.\footnote{On dimensional grounds, for terms that are at most
log-divergent, $r$ can only appear in terms that go as $\mathcal{O}(x/r)$.
To preserve the symmetries of the space such terms can only appear in the metric (or its inverse or the volume element for the space).
By contrast, a term that can have $1 / \e$ divergence can also have terms 
of order $(\log \e) / r^2$.}

We take the flat limit after performing similar coordinate transformations as in Appendix A,
most significantly, the transformation $z' \rightarrow \tilde{z}$ that sets $1 + \sigma^2(z',\alpha w') = (1+\vert\tilde{z}\vert^2)(1+\e)$.
Thus, in the $r \rightarrow \infty$ limit, we can take the regulated propagator to be
\beq
\tilde{G}_{\rm Reg} (x,y) \rightarrow {1 \over 2(\vert x-y\vert^2 +\e)}
\label{B1}
\eeq
Here we only included the first term in the propagator, as in the 
$r \rightarrow \infty$ limit it is the only term that survives
for both the scalar and $Y=-2$ fields.
The IR regulator doesn't show up in UV divergent terms. 

In $\Gamma_0$ the only term that can have $1 / \e$ divergence
 is the WZW action which happens to have a finite coefficient (see above). 
For log-divergent terms we expand $\la {\hat J} \ra$ further
in $\nabla H H^{-1}$.
However, since we calculate them in the limit $r \rightarrow \infty$,
the result will be the same for scalar and $Y=-2$ fields. Thus we can use our
result from \cite{KMN1}, where we calculated the log-divergent term
for the scalar current. Here, we simply need to double it, as there is one
term coming from the scalar part of $\Gamma_0$ and one from the $Y=-2$ part.
\beq
\Gamma_0 ={\log \e \over 12} \int \, \Tr \, (\bnabla (\nabla H H^{-1}))^2 + {\rm finite}
\label{B2}
\eeq

For $\Delta \Gamma$ we can take $M \rightarrow H$ and $M^{\dagger} \rightarrow 1$, as
discussed in section 6.2. Thus,
\beqar
\Delta \Gamma &=& \la S_1 \ra - {1 \over 2!} \la S_1^2 \ra + {1 \over3!} \la S_1^3 \ra + \cdots \nonumber\\
&=& \Delta \Gamma^{(1)} + \Delta \Gamma^{(2)} + \Delta \Gamma^{(3)} + \cdot
\label{B3}
\eeqar
where
\beqar
S_1 &=& \int \, d\mu \Bigl[ \bar{C}^{\a} (HaH^{-1} \cdot \ba)^{\a \b}C^{\b} + C^\a (\ba \cdot \D)^{\a \b} C^\b + \bar{C}^\a (-H a H^{-1} \cdot \bnabla )^{\a \b} \bar{C}^\b  \nonumber\\
&& \hskip .4in  + C^\a ( - \e^{\bar{i}\bar{j}}\ba_{\bar{i}} \bnabla_{\bar{j}})^{\a \b} B^\b + \bar{C}^\a (\e^{ij} H a_i H^{-1} \nabla_j )^{\a \b} \bar{B}^\b \Bigr]
\label{B4}
\eeqar
and $\la C^{\a} \bar{C}^{\b} \ra = \G^{\rm scalar} = (-\bnabla \cdot \D)^{-1}_{\rm scalar}$ and $\la \bar{B}^\a B^\b \ra = \tilde{\G}^{\rm Y=-2} = (-\bnabla \cdot \D)^{-1}_{\rm Y=-2}$.
In carrying out various calculations, we will be expanding 
$\G^{\rm scalar}$ and $\tilde{\G}^{\rm Y=-2} $ in terms of the corresponding 
free propagators $G$ and $\tilde{G}$, 
as in (\ref{cpII-53}).
For $Y = -2$,
the UV regularized form of the free propagator is given in
(\ref{cpII-45}), (\ref{cpII-48}).
The scalar propagator was given in \cite{KMN1} and has the form
\beq
G(z,y) = {1\over 2 \, s}   - {1\over 2} \log\left( {s \over 1+s}\right)
- {3\over 4}, \hskip .2in s = \sigma^2_{z,y}
\label{B4a}
\eeq
with the replacement of $s$ by $\sigma^2 (Z, Y')$ to take account
of regularization.

For the mass term, we keep $r$ finite and perform similar calculations as in Appendix A. The log-divergent terms can be obtained by calculating in the flat space limit, and then upgrading the metric and volume factors to the curved space expressions.
Then, expanding each $\G$ in $\nabla H H^{-1}$, we find the following UV divergent terms
\beqar
\Delta \Gamma^{(1)} &=& \la S_1 \ra = \int \Tr \, H a H^{-1} \ba \, \G^{\rm scalar}_{\rm Reg} \nonumber\\
&=& \left({1\over 2\e} - {1\over 2 r^2} \log \e \right) \int \Tr \, H a H^{-1} \ba + {\log \e \over 4} \int \Tr \, HaH^{-1}\ba \bnabla (\nabla H H^{-1}) 
\label{B5}\\
\Delta \Gamma^{(2)} &=& - {1 \over 2!} \la S_1^2 \ra  \nonumber\\
&=& - {1 \over 2} \int \Tr \Bigl[HaH^{-1} \ba \, \G^{\rm scalar} \, HaH^{-1} \ba \, \G^{\rm scalar}_{\rm Reg}  \nonumber\\
&& \hskip .5in - 4 \ba \cdot \D \G^{\rm scalar} \, H a H^{-1} \cdot \bnabla \G'^{\rm scalar}_{\rm Reg} \nonumber\\
&&\hskip .5in  + 2 \e^{\bar{i} \bar{j}} \bnabla_{\bar{i}} (\ba_{\bar{j}} \G^{\rm scalar} ) \, \e^{ij} Ha_i H^{-1} \D_j \tilde{\G}^{\rm Y=-2}_{\rm Reg} \Bigr] \nonumber\\
&=& \left( - {3 \over 4 \e} + {1 \over  r^2} \log \e \right) \int \Tr \, H a H^{-1} \ba \nonumber\\
&& + \log \e \int \Tr \, \left[ {1\over 4} (H a H^{-1} \ba)^2 - {5 \over 12} Ha H^{-1}\ba \bnabla (\nabla H H^{-1}) \right.\nonumber\\
&& \hskip 1in -{1\over 12} g^{a \ba} g^{b \bar{b}} \bnabla_{\ba} \ba_{\bar{b}} \D_a (H a_b H^{-1}) \nonumber\\
&& \hskip 1in  \left. -{1\over 12} g^{a \ba} g^{b \bar{b}} \bnabla_{\bar{b}} (\nabla_a H H^{-1} ) [ \ba_{\ba}, H a_b H^{-1} ] \right] 
\label{B6}\\
\Delta \Gamma^{(3)} &=& {1 \over 3!} \la S_1^3 \ra \nonumber\\
&=& {1 \over 3} \int \Tr \Bigl[ HaH^{-1} \ba \, \G^{\rm scalar} \, HaH^{-1} \ba \, \G^{\rm scalar} \, HaH^{-1} \ba \, \G^{\rm scalar}_{\rm Reg}  \nonumber\\
&& \hskip .4in -12 \ba \cdot \D \G^{\rm scalar} \, H a H^{-1} \ba \, \G^{\rm scalar} \, Ha H^{-1} \cdot \bnabla \G'^{\rm scalar}_{\rm Reg} \nonumber\\
&& \hskip .4in  +3 \e^{\bar{i}\bar{j}}\bnabla_{\bar{i}} ( \ba_{\bar{j}} \G^{\rm scalar} ) \,  Ha H^{-1} \ba \, \G^{\rm scalar} \, \e^{ij} H a_i H^{-1} \D_j \tilde{\G}^{\rm Y=-2}_{\rm Reg} \Bigr] \nonumber\\
&=& - {5 \over 4} \log \e \int \Tr \, (HaH^{-1} \ba)^2 
\label{B7}\\
\Delta \Gamma^{(4)} &=& - {1 \over 4!} \la S_1^4 \ra \nonumber\\
&=& - {1 \over 4} \int \Tr \Bigl[ 16 \ba \cdot \D \G^{\rm scalar} \, HaH^{-1} \cdot \bnabla \, \G'^{\rm scalar} \ba \cdot \D \, \G^{\rm scalar} \, HaH^{-1} \cdot \bnabla \G'^{\rm scalar}_{\rm Reg}  \nonumber\\
&& \hskip .5in - 16 HaH^{-1}\cdot \bnabla \G'^{\rm scalar} \, \ba \cdot \D \, \G^{\rm scalar} \, \e^{ij} H a_i H^{-1} \D_j \tilde{\G}^{\rm Y=-2} \, \e^{\bar{i}\bar{j}} \bnabla_{\bar{i}} (\ba_{\bar{j}} \G^{\rm scalar}_{\rm Reg} ) \nonumber\\
&& \hskip .5in +2 \e^{\bar{i}\bar{j}}\bnabla_{\bar{i}} (\ba_{\bar{j}} \G^{\rm scalar} ) \, \e^{ij} Ha_i H^{-1} \D_j \tilde{\G}^{\rm Y=-2} \e^{\bar{k}\bar{l}} \bnabla_{\bar{k}} (\ba_{\bar{l}} \G^{\rm scalar} ) \, \e^{kl} H a_k H^{-1} \D_l \tilde{\G}^{\rm Y=-2}_{\rm Reg} \Bigr] \nonumber\\
&=& {13 \over 12} \log \e \int \Tr \, (HaH^{-1} \ba)^2
\label{B8}
\eeqar
where in the above $\G' = (-\D \cdot \bnabla)^{-1}$ and the traces are in the adjoint representation.

Combining the four terms
\beqar
\Delta \Gamma &=& \left( -{1 \over 4 \e} +{1 \over 2 r^2} \log \e \right) \int \Tr \, H a H^{-1} \ba \nonumber\\
&& + {1 \over 12} \log \e \int \Tr \Bigl[ [\ba, HaH^{-1} ] \bnabla (\nabla H H^{-1}) + (HaH^{-1} \ba)^2  
\label{B9}\\
&& \hskip .9in -g^{a \ba} g^{b \bar{b}} \Bigl( \bnabla_{\ba} \ba_{\bar{b}} \D_a (H a_b H^{-1}) + \bnabla_{\bar{b}}(\nabla_a H H^{-1}) [\ba_{\ba}, Ha_b H^{-1}] \Bigr) \Bigr] \nonumber
\eeqar

Putting together $\Gamma_0$ and $\Delta \Gamma$ from (\ref{B2}) and (\ref{B9}) above
\beqar
\Gamma &=& \Gamma_0 + \Delta \Gamma \nonumber\\
&=& \left( -{1 \over 4 \e} +{1 \over 2 r^2} \log \e \right) \int \Tr \, \ba H a H^{-1} \nonumber\\
&& + {1 \over 12} \log \e \int \Tr \Bigl[ (\bnabla(\nabla H H^{-1}))^2 + [\ba, HaH^{-1} ] \bnabla (\nabla H H^{-1}) + (\ba HaH^{-1} )^2 
\nonumber\\
&& \hskip .9in -g^{a \ba} g^{b \bar{b}} \Bigl( \bnabla_{\bar{b}}(\nabla_a H H^{-1}) [\ba_{\ba}, Ha_b H^{-1}] + \bnabla_{\ba} \ba_{\bar{b}} \D_a (H a_b H^{-1}) \Bigr) \Bigr]
\label{B10}
\eeqar
which are the results for the mass term and log-divergent terms in (\ref{mass3}) and (\ref{log1}).

%%%%%%%%%%%%%%%%%%%%%%%%%%%%%%%%%%%%%%%%%%%%%%%%
%%%%%%%%%%%%%%%%%%%%%%%%%%%%%%%%%%%%%%%%%%%%%%%%

%%%%%%%%%%%%%%%%%%%%%%%%%%%%%%%%%%%%%%%%%%%%%%%%
%%%%%%%%%%%%%%%%%%%%%%%%%%%%%%%%%%%%%%%%%%%%%%%%
%%%%%%%%%%%%%%%%%%%%%%%%%%%%%%%%%%%%%%%%%%%%%%%%
%%%%%%%%%%%%%%%%%%%%%%%%%%%%%%%%%%%%%%%%%%%%%%%%
\end{document}